%% file: paper.tex
\documentclass{sig-alternate-10pt}

\usepackage{times}
\usepackage[scaled=0.92]{helvet}
\usepackage{indentfirst}
\usepackage[T1]{fontenc}
\usepackage{textcomp}
\usepackage[table]{xcolor}
\usepackage{graphicx}
\usepackage{caption}
\DeclareCaptionType{copyrightbox}
\usepackage{subcaption}
\usepackage[
bookmarks=false,
breaklinks=true,
colorlinks=true,
linkcolor=black,
citecolor=blue,
urlcolor=black,
pdfpagelayout=SinglePage
]{hyperref}
\usepackage[all]{hypcap}
\usepackage[hyphenbreaks]{breakurl}
\usepackage{url}
\usepackage{mathptmx}
\usepackage{amsmath,amssymb}
\usepackage{multirow}
\usepackage[noadjust]{cite}
\usepackage[scaled]{couriers}
\usepackage{balance}
\usepackage[smallbib,obey]{fixacm}
\usepackage[linesnumbered,ruled,vlined]{algorithm2e}
\SetAlgoSkip{}
\usepackage{microtype}
\usepackage{epsfig}
\usepackage{appendix}

\newdef{definition}{Definition}

\newcommand{\squishlist}{
 \begin{list}{${\bullet}$}
  { \setlength{\itemsep}{0pt}
     \setlength{\parsep}{3pt}
     \setlength{\topsep}{3pt}
     \setlength{\partopsep}{0pt}
     \setlength{\leftmargin}{1.5em}
     \setlength{\labelwidth}{1em}
     \setlength{\labelsep}{0.5em} } }

\newcommand{\squishend}{
  \end{list}  }

\begin{document}
%\setcopyright{none}
%\doi{ }
%\isbn{ }

\makeatletter
\def\@copyrightspace{\relax}
\makeatother

\title{TimeWeaver: Opportunistic One Way Delay Measurement via NTP}
\date{}

\newcommand{\superscript}[1]{\ensuremath{^{\textrm{#1}}}}
\def\sharedaffiliation{\end{tabular}\newline\begin{tabular}{c}}
\def\wu{\superscript{*}}
\def\wp{\superscript{+}}
\def\wg{\superscript{\dag}}
\def\wk{\superscript{\#}}

\numberofauthors{1}
\author{
  \alignauthor Ramakrishnan Durairajan\wg, Sathiya Kumaran Mani\wk, Paul Barford\wk\wp, Rob Nowak\wk, Joel Sommers\wu \\
  \sharedaffiliation
  \begin{tabular}{cccc}
    \affaddr{{\wg}University of Oregon{\ }} &  \affaddr{{\wk}University of Wisconsin - Madison{\ }} & \affaddr{{\wp}comScore, Inc. {\ }} & \affaddr{{\wu}Colgate University{\ }} \\
  \end{tabular}
}

\maketitle

\sloppypar

\begin{abstract} \input{abstract} \end{abstract}

\maketitle
%%%% \vspace{-0.2cm}
\section{Introduction} \label{sec:introduction} \input{introduction_pb}
%%%% \vspace{-0.2cm}
\section{Background} \label{sec:background} \input{background}
%%%% \vspace{-0.2cm}
\section{NTP Data Collection} \label{sec:datasets} \input{datasets}
%%%% \vspace{-0.2cm}
\section{Extracting One Way Delays from NTP data} \label{sec:filtering} \input{filtering}
%%%% \vspace{-0.2cm}
\section{Characteristics of OWDs} \label{sec:characteristics} \input{characteristics}
\section{Internet distance estimation from One Way Delays} \label{sec:results} \input{matrix}
\section{Additional applications} \label{sec:application} \input{applications}
%%%% \vspace{-0.2cm}
\section{Related Work} \label{sec:relatedwork} \input{related}
%%%% \vspace{-0.2cm}
\section{Summary and Future Work} \label{sec:summary} \input{summary}
\balance
\bibliographystyle{IEEEtran}
\bibliography{paper}
% \appendix
% \section{Summary of NTP server logs used in TimeWeaver} \label{app:ntpDataSummary} \input{bigtable}
\end{document}

%% file: abstract.tex
One-way delay (OWD) between end hosts has important implications for Internet applications, protocols, and measurement-based analyses. We describe a new approach for identifying OWDs via passive measurement of Network Time Protocol (NTP) traffic. NTP traffic offers the opportunity to measure OWDs accurately and continuously from hosts throughout the Internet. Based on detailed examination of NTP implementations and in-situ behavior, we develop an analysis tool that we call TimeWeaver, which enables assessment of precision and accuracy of OWD measurements from NTP. We apply TimeWeaver to a $\sim$1TB corpus of NTP traffic collected from 19 servers located in the US and report on the characteristics of hosts and their associated OWDs, which we classify in a precision/accuracy hierarchy. To demonstrate the utility of these measurements, we apply iterative hard-threshold singular value decomposition to estimate OWDs between arbitrary hosts from the highest tier in the hierarchy. We show that this approach results in highly accurate estimates of OWDs, with average error rates on the order of less than 2\%. Finally, we outline a number of applications---in particular, IP geolocation, network operations and management---for hosts in lower tiers of the precision hierarchy that can benefit from TimeWeaver, offering directions for future work.

%% file: introduction_pb.tex
%%%% \vspace{-0.2cm}
End-to-end network latency plays a fundamental role in behavior and performance at different layers of the protocol stack.  As such, measurements of network latency are included in many protocols and systems in order to adjust their behavior to network conditions.  A canonical example is TCP's measurement of round-trip time (RTT) to adjust its sending behavior.  Additionally, numerous network-focused applications have been developed over the years that are based specifically on latency measurements including network positioning and distance estimation ({\em e.g.,}~\cite{dabek2004,mao2006ides,chen2009phoenix,liao2010network}), available bandwidth estimation ({\em e.g.,}~\cite{jain2002end,sommers2006proposed})) and IP geolocation ({\em e.g.,}~\cite{Gueye06,Wong07})) to name a few.

Latency depends intrinsically on the routes that packets traverse between end points.  If routes between hosts are symmetric, then RTT measurements would be appropriate for many protocols and applications.  RTT measurements would also be a reasonable proxy for understanding network proximity, {\em e.g.,} in the context of selecting a ``closest server".  Indeed, this notion of latency-based proximity estimation is the basis for standard client redirection in content delivery networks (CDNs)~\cite{krishnan2009moving} and certain DHTs ({\em e.g.,}~\cite{rowstron2001pastry}).  However, routes are frequently asymmetric~\cite{paxsonthesis,sanchez13}, which reduces the utility of RTT measurement for proximity estimation and for real-time applications ({\em e.g.,} streaming and gaming), and makes one way delay (OWD) an important measurement objective. 

Measuring end-to-end latency (RTT or OWD) has several basic challenges.  First, there are a number of factors that contribute to latency in addition to the physical path including queuing and processing delays by nodes in the network.   While one can consider the physical path to be relatively stable over moderate timescales, the latter two factors can introduce variability into measurements on shorter timescales~\cite{zpds01}.  This implies the need to specify a target measurement metric, {\em e.g.,} average RTT or minimum OWD (minOWD), and to devise an appropriate methodology for collecting and analyzing measurement data.  Central to measurement methodology design are the precision and accuracy requirements.  These may imply a relatively simple measurement system or a complex infrastructure based on dedicated hardware.  Finally, measuring OWD has the additional requirement that host clocks must be synchronized.

In this paper, we consider the problem of measuring OWDs in the Internet.  Our specific interest is in measuring OWDs at scale ({\em i.e.,} from many hosts in the Internet), with high precision and accuracy
and without the need for complex, dedicated systems.  We posit that such measurements could be applied to a wide variety of timely and important problems including those mentioned above.

The basis for our work is the vast quantity of OWD measurements that can be extracted from Network Time Protocol (NTP) packet exchanges---both from client to server, and server to client.  NTP is pervasively used by hosts in the Internet to synchronize their clocks with high fidelity time sources.  An intrinsic component of the protocol is estimation of OWD, which is used in the client clock adjustment algorithm.

We begin by developing a method for extracting accurate OWDs from NTP data.  As pointed out in~\cite{Durairajan2015HotNets}, OWD measurements extracted from NTP packets are not always accurate.  Our filtering method is based on a detailed analysis of the NTP codebase~\cite{ntpCodeBase}.  Our analysis reveals regimes in the NTP packet exchange process that are observable in the traces, and which can be used to infer strong synchronization between clocks and thus accurate OWD measurement.  We realize our filtering method in a framework that we call TimeWeaver, which organizes OWDs into measurement precision tiers that are based on observed client behavior.  The tiers provide context for understanding the accuracy and utility of the measurements.

We assembled a $\sim$1TB corpus of NTP packet data, which was collected from 19 US-based NTP servers over a period of 30 days\footnote{Our study can be repeated at other locations by collecting NTP packet traces, which can be easily be obtained from NTP administrators (see \S\ref{sec:discussion}).}. There are 162,798,893 IPv4 and 6,056,609 IPv6 unique client addresses evident in the data.  Examination of the client IP addresses indicates hosts from around the world are configured to synchronize to these servers.  

We apply TimeWeaver to our NTP data and assess the resulting OWDs vs. a prior filtering method and vs. active probe-based RTT measurements.  We find that the sum of forward and reverse paths OWDs from the high precision/accuracy tier correlate well with probe-based measurements, and that TimeWeaver filters with much higher accuracy than~\cite{Durairajan2015HotNets}.  Our analysis of OWD measurements reveals diverse characteristics including a range of OWDs in all tiers but the range is more narrow (typically under 100ms) in the high precision/accuracy tier.  
% Our analysis of OWD measurements reveals diverse characteristics.  We find a range of OWDs in all tiers but the range is more narrow (typically under 100ms) in the high precision/accuracy tier.  We also find that a large fraction of the paths between client and server are asymmetric, indicating that RTT/2 is a risky approximation for OWD. 

Next, we further demonstrate the utility of OWD estimates based on opportunistic NTP measurements by applying them to the problem of inter-host distance estimation.  While prior studies have considered this problem ({\em e.g.,}~\cite{Francis2001,ng2002}), our formulation differs in that we consider minOWDs (as opposed to round-trip times) in the context of a Euclidean space. A potentially significant issue in prior distance estimation methods is triangle inequality violations (TIV) caused by round-trip routes. We argue that our use of minOWD diminishes the susceptibility to TIV.  Our argument is supported by comparing error rates on distance estimates produced by our method versus those produced by a variety of prior methods including DMF, which was designed to be immune to TIV~\cite{chen2009phoenix}.

Our matrix completion method for estimating inter-host distance is based on iterative hard-threshold singular value decomposition~\cite{chunikhina2014}.  This algorithm iterates between truncating the SVD of the current estimate to a user-specified rank $r$, and then replaces the values in the observed entries with their original (observed) values.  For Euclidean space, we consider $r=4$ and apply the algorithm to minOWD measurements from NTP clients that contact more than one server.  We find that the resulting distance estimates are highly accurate, with errors on the order of about 2\%.

In summary, this paper makes the following contributions. First, we develop a filtering algorithm that we implement in a tool called TimeWeaver, which enables accurate extraction of OWDs from NTP packets.  Second, we report OWD characteristics of Internet hosts by applying our filtering algorithm to a $\sim$1TB corpus of NTP trace data. Our analysis highlights %the prevalence of asymmetric routing and 
the characteristics of OWDs in distinct precision tiers. Third, we describe a new SVD-based method for inter-host distance estimation. We show that when this method is applied to minOWD measures from NTP, resulting distance estimates are highly accurate. Overall, we show that TimeWeaver applied to passive measurements of NTP produces highly accurate OWD estimates that could be applied in a wide variety of applications.
%These benefits emerge directly from passive measurement of NTP.

%% file: background.tex
%%%% \vspace{-0.2cm}
Synchronizing independently running clocks and achieving temporal consistency in a distributed system is a compelling and challenging problem. The most widely used protocol in the Internet for time synchronization is the Network Time Protocol (NTP). The first specification for NTP appeared in 1985 as RFC 958~\cite{rfc958}. The current recommended version of the three-decade-old protocol is NTPv4~\cite{rfc5905}, which is backward-compatible with the most widely implemented version in the Internet, NTPv3~\cite{rfc1305}.

The NTP ecosystem is composed of a hierarchy of servers.  Starting at the top-level are servers with high-precision time sources such as GPS-based and atomic clocks.  These servers, referred to as stratum 0, offer high-quality timing information to servers in the next level, stratum 1, which are also known to as \textit{primary} servers. Stratum 2 or \textit{secondary} servers connect to stratum 1 servers, etc., all the way down to stratum 15.  In addition to connecting to a source up the hierarchy, NTP servers may also peer with others at the same level for redundancy.

Hosts in the Internet typically synchronize time with more than one server in order to compute a precise time estimate. Even though a host running a commodity operating system is configured with default NTP server(s) to synchronize time ({\em e.g.}, \url{time.windows.com}, \url{time.apple.com}, \url{0.pool.ntp.org}), it can be manually configured to use a specific NTP server or a set of servers.  Recent efforts like {\tt ntp.org} also maintain lists of stratum 1 and stratum 2 servers that can be used after acquiring permission from the server administrators.  NTP hosts or clients typically connect to reference clocks that are stratum 2 or higher.

Every NTP client host runs the {\tt ntpd} daemon, which in turn runs several filtering algorithms and heuristics to synchronize its clock with reference clocks in the Internet.  At a high level, {\tt ntpd} operates by exchanging timestamps with its reference servers (in a process called {\em polling}).  When and how often reference servers are polled is governed by the \textit{clock discipline algorithm}~\cite{ntpClockDiscipline}. In most operating systems, the polling interval starts with {\tt minpoll} (64s) intervals and may eventually increase in steps to {\tt maxpoll} (1024s) intervals.  As part of its operation, the algorithm measures round-trip delay, jitter and oscillator frequency wander to determine the best polling interval~\cite{nptfaq}. 

Four timestamps are included in NTP packets as a result of each polling round: the time at which a polling request is sent ($t_0$), the time at which the request is received at the server ($t_1$), the time at which the response is sent by the server ($t_2$), and the time at which the response is received by the client ($t_3$).  These timestamps are not set until after the completion of a handshake between client and server, which is indicated by the inclusion of an IPv4 address or hash of an IPv6 address in the {\tt ref id} field of a request packet.  Unfortunately, the logs do not contain explicit information regarding whether a client's clock is in ``good'' synchronization with the server, which is when differences between timestamps would offer the most accurate indication of OWD. In addition, a sizable number of hosts in the Internet use Simple NTP (SNTP)~\cite{rfc1769}, which sets all packet fields to zero except the first octet which mainly contains metadata ({\em e.g.}, version number, stratum, poll interval, etc.).  Due to the lack of explicit synchronization information in the NTP packets and the prevalence of SNTP clients ({\em e.g.}, mobile and wireless hosts)~\cite{sathiya2016}, we develop a framework to classify the precision of timestamps in an NTP packet as we discuss in 
%% one of four of varying precisions as we discuss in 
\S\ref{sec:filtering}.

%% file: datasets.tex
%% 9 cities -- Atlanta, Chicago, Edison, Jackson, Madison, Monticello, Salt Lake City, Urbana-Champaign, Philadelphia
%% 19 servers -- 5 stratum-1, 14 stratum-2 servers
%% different entities -- 2 Independent, 7 ISP, 3 Private, 7 University
%%In this section, we describe the data sets collected from 9 different cities in the U.S. and provide high-level statistics of the 19 NTP servers used in our study.
%IPv4 total:  162,798,893
%IPv6 total:  6,056,609
%%%% \vspace{-0.2cm}
In this section, we describe the datasets collected from 19 NTP servers located in 9 different cities in the U.S., which are the focus of our study.

%% has the big statistics table
\input{bigtable}

%%%% \vspace{-0.2cm}
\subsection{NTP data collection}
%%%% \vspace{-0.2cm}

To collect the NTP log data used in our study, we contacted several NTP administrators and explained our research goals. Eight administrators responded by providing datasets in the form of full pcap (``tcpdump'') traces from a total of 19 NTP servers. The servers include a combination of {\em (1)} 7 NTP servers in Chicago (IL), Edison (NJ), and Salt Lake City (UT) from 2 different Internet service providers, {\em (2)} 3 commercial NTP servers in Jackson (WI) and Monticello (IA), {\em (3)} 7 university campus NTP servers in Madison (WI) and Urbana-Champaign (IL), and {\em (4)} 2 independent/community NTP servers in Atlanta (GA) and Philadelphia (PA). To facilitate network latency analysis, we developed a lightweight tool (about 800 lines of C code) to process/analyze the NTP logs.\footnote{The tool and the datasets used in this study will be openly available to the community upon publication.}

Our efforts to amass server logs from NTP administrators can be replicated by anyone in the community due to the ubiquity of NTP servers in the Internet ({\em e.g.,} there are over 3.6K servers in {\tt pool.ntp.org} alone~\cite{ntppool}). When we began our study, we %created a portal ~\cite{ntpStudy} explaining our research goals and
reached out directly to 12 NTP server administrators.  All the administrators responded positively and were willing to help with access to the server logs. Out of these 12 willing administrators, we eventually collected data from 8 of them due to problems such as modification of internal policies to facilitate data collection, server unavailability, server relocation due to attacks, etc. at 4 of the sites.  Still, this represents a 65\% success rate with little effort in obtaining NTP packet traces from server administrators, which we find quite encouraging. 

%%%% \vspace{-0.2cm}
\subsection{Basic statistics}
\label{sec:discussion}
%%%% \vspace{-0.2cm}

Table~\ref{tab:statsBasic} summarizes the basic statistics from each of the NTP server logs and some of the key properties of the servers such as stratum, IP version supported, number of measurements observed in the log files, and number of clients. The selection includes 5 stratum-1 servers and 14 stratum-2 servers with a combination of both IPv4 and IPv6 support. These logs include a total of 6,369,784,837 latency measurements to 162,798,893 IPv4 and 6,056,609 IPv6 worldwide clients, as indicated by unique IP addresses, collected over a period of one month from Nov., 2015 to Dec., 2015. % From these 6,339,977,365 latency measurements, we filtered 5,612,442,552 measurements due to malformed headers, packet errors, missing timestamps, negative latency values, and handshake packets, leaving us with 727,534,813 latency measurements which form the basis of our analysis.  While these measurements indicate that we remove a significant portion of NTP protocol exchanges, the remaining usable packets still contain latency information collected from a significant and broad slice of the Internet, which we argue would require a great deal of effort to collect using other means, {\em e.g.}, active measurements.

%% file: bigtable.tex
\begin{table*}[!htbp]
\centering
  {\caption{\label{tab:statsBasic}{{\bf Summary of NTP server logs used in this study.}}}}
\begin{tabular}{ |c|c|c|c|c|c|c|c| }
\hline
Server							&	Server	&	Server	&	IP		&	Server 		& 	Total			&    Total			& 	Total			\\
Location							&	ID		&	Stratum	&	Version	&	Organization 	& 	Measurements	&    IPv4			& 	IPv6			\\
								&			&			&			&				&				&    Clients		& 	Clients		\\ \hline \hline
Atlanta, GA						& 	AG1		&	2		& 	v4		&	Independent  	&	349,917,829	&    12,889,722		& 	0			\\ \hline
\multirow{4}{*}{Chicago, IL}			&	CI1		&	2		&	v4/v6	&	ISP  			&	23,201,076	&    1,337			& 	88			\\
								&	CI2		&	2		&	v4/v6	&	ISP  			&	22,163,583	&    1,009 			&	71			\\
								&	CI3		&	2		&	v4/v6	&	ISP  			&	24,836,284	&    701			&	62			\\
								&	CI4		&	2		&	v4/v6	&	ISP  			&	23,846,973	&    573 			&	44			\\ \hline
\multirow{2}{*}{Edison, NJ}			&	EN1		& 	2		&	v4/v6	&	ISP  			&	13,166,749	&    581			&	46			\\
								&	EN2		& 	2		&	v4/v6	&	ISP  			&	13,381,258	&    536			&	41			\\ \hline
\multirow{2}{*}{Jackson, WI}      			&	JW1		& 	1		& 	v4		& 	Commercial  	&	11,498,989	&    337,015		&	0			\\
								&	JW2		& 	1		& 	v4		& 	Commercial  	&	40,330,009	&    864,845		&	0			\\ \hline
\multirow{4}{*}{Madison, WI}			&	MW1		& 	1		&	v4		&	University  	&	5,451,294		&    20,589		&	0			\\
								&	MW2		& 	2		&	v4		&	University  	&	1,850,765,317	&    60,682,989		&	0			\\
								&	MW3		& 	2		&	v4		&	University  	&	386,487,947	&    26,997,177		&	0			\\
								&	MW4		& 	2		&	v4		&	University  	&	355,913,460	&    16,758,046		&      	0			\\ \hline
Monticello, IA						& 	MI1		& 	1		& 	v4		& 	Commercial  	&	1,899,642,404	&    27,133,385		&	0			\\ \hline
Philadelphia, PA					& 	PP1		& 	2		& 	v4/v6	& 	Independent	&	10,090,072	&    690,486		&	0			\\ \hline
Salt Lake City, UT					& 	SU1		& 	1		& 	v4/v6	& 	ISP  			&	590,431,652	&    16,206,848		&      6,052,784  	\\ \hline
\multirow{3}{*}{Urbana-Champaign, IL}	&	UI1		&	2		&	v4/v6	&	University  	&	302,622,909	&    58,967		&	1,363		\\
								&	UI2		&	2		&	v4/v6	&	University  	&	270,990,678	&    98,159		&	1,147		\\ 
								&	UI3		&	2		&	v4/v6	&	University  	&	175,046,354	&    55,928		&	963			\\ \hline
\end{tabular}
\end{table*}

%IPv4 total:  162,798,893
%IPv6 total:  6,056,609

%% file: filtering.tex
%%%% \vspace{-0.2cm}
In this section, we describe a framework called TimeWeaver for extracting and classifying OWDs from NTP packets.  Our examination of NTP traces along with observations of others ({\em e.g.},~\cite{ntpFilter, Ridoux10, Durairajan2015HotNets}) imply that latency measurements available through NTP packet exchanges may be skewed.  
%% Moreover, there may be inherent limitations due to the way in which NTP is designed to work. 
A key aspect of the TimeWeaver framework is that in addition to adjusting NTP-derived latency measurements for skew, we assign measurements into different \textit{precision tiers}\footnote{Our original goal was a quantitative precision framework based on standard deviations of repeated measurements.  However, the highly dynamic nature of NTP renders this approach unreliable. We will show that our tiered framework provides a reliable and useful context for interpreting the OWDs extracted by TimeWeaver.} as we discuss below.

%\subsection{Filtering NTP packet traces}
%%%% \vspace{-0.2cm}
\subsection{OWD precision framework}
%%%% \vspace{-0.2cm}

{\bf Overview.} The TimeWeaver framework adopts the notion of precision discussed by Paxson~\cite{Paxson04}, specifically that it is ``the maximum exactness that a tool's design permits.''  Thus, the basic assumption we start with is that our precision assignment framework must be {\em NTP-specific}.  That is, given information available within the NTP packet traces ({\em e.g}, timestamps relative to client and server clocks, polling values), we do not expect to have success with a na\"ive approach like excluding extreme OWD values, or by only including values close to the minimum observed OWD.  The reason, again, is that there is no meta-information available in protocol messages to indicate whether a client has reached good/close synchronization with the server. Instead, our approach is explicitly designed to exploit the ways in which the protocol {\em behaves} in response to good synchronization or events that degrade synchronization to create \textit{tiers of precision}, each of which is suitable for various applications of interest. Extracted OWDs are assigned to a specific tier based on the inferred level of synchronization and the number and quality of measurements. Specifically, we define the following four precision tiers:

\squishlist
\item {\em Tier 0}: These samples are from SNTP/NTP clients issuing a one-shot synchronization request. Unfortunately, no OWD information is available in these samples.
\item {\em Tier 1}: This tier includes OWD measurements derived from clients using NTP which often exchange multiple packets with servers. The clients are either moving towards or away from close synchronization with the servers and the OWDs extracted are typically greater than one second with respect to the reference.
\item {\em Tier 2}: Similar to Tier 1, OWD measurements in their tier are from clients that exchange multiple packets with servers and cannot be confirmed to be in close synchronization.  The main difference with Tier 1 is that the OWDs are less than one second.\footnote{We set the threshold to one second similar to prior efforts ({\em e.g.}, iPlane~\cite{Madhyastha2006} and Hubble~\cite{Katz-Bassett2008}  sets a two-second timeout for RTT-based probes, and RIPE Atlas~\cite{ripeatlas} sets a timeout of one second for their ICMP echo requests~\cite{mailPhilip}).}
\item {\em Tier 3}: This tier includes highly accurate OWD measurements from clients which are observed to be tightly synchronized with their NTP references.
\squishend

%We start with the basic assumption that our filtering approach must be {\em NTP-specific}. That is, given information available within the NTP packet traces ({\em i.e.}, timestamps relative to client and server clocks, plus some additional information), we do not expect to have success with a na\"ive approach like excluding extreme OWD values, or by only including values close to the minimum observed OWD.  The reason, again, is that there is no meta-information available in protocol messages to indicate whether a client has reached good/close synchronization with the server.  Instead, our method is explicitly designed to exploit the ways in which the protocol {\em behaves} in response to good synchronization, or events that degrade synchronization.

We first exploit NTP behavior by considering the polling operation of clients, dividing them into two basic classes:  constant or non-constant polling. Our motivation is similar to that of prior work~\cite{Durairajan2015HotNets} in that we attempt to take advantage of polling behavior in order to detect whether the client is in good or poor synchronization with the server.  For example, an intended protocol behavior is for a client to increase its polling rate (reduce the polling interval) in response to poor synchronization. Likewise, in response to detection of good synchronization, a client may reduce its polling rate (increase the polling interval). Unfortunately, this is not sufficient, as there are clients that do not vary their polling rate at all.  For these clients, we use similar heuristics to those within the NTP protocol~\cite{nptfaq} and code~\cite{ntpCodeBase} to identify high-quality latency samples as we discuss below.

{\bf Algorithm.} The key steps of TimeWeaver's precision assignment algorithm are shown in Algorithm~\ref{alg:filtering}.  As part of developing this algorithm, we conducted a detailed study of the NTP codebase~\cite{ntpCodeBase}, request and response transactions, protocol behaviors, packet fields and packet selection heuristics.  We also experimented with different NTP client and server configurations ({\em e.g.}, Mac OS, Linux, and Windows) in a controlled laboratory setting.  Our goal was to understand the {\em operational aspects} of NTP in detail.  Specifically, we conducted measurements in different settings: {\em (i)} distant client synchronizing with a local NTP server, {\em (ii)} local client synchronizing with a distant NTP server and {\em (iii)} local client synchronizing with a local server. From our source code analysis and controlled experiments we identified two specific features of the protocol to leverage in our filtering algorithm: the client-estimated ground truth RTT (gtRTT) value, which is used when a client polls at a {\em constant} rate, and the jiggle counter heuristic~\cite{nptfaq} used in NTP's client selection algorithm, which is used when a client exhibits {\em non-constant} polling behavior. In addition to these NTP-specific filtering methods, we found that we needed to eliminate spikes in OWD samples, as we discuss below.

{\bf Non-constant polling behavior.} For clients with non-constant polling values, we use insights from the client selection heuristic~\cite{nptfaq} to select what are likely to be high-quality latency samples.  For a given polling value ($2^{P^{c}_{e}}$, where $P^{c}_{e}$ is the polling exponent), the algorithm requires at least $N$ samples (where N = 30/${P^{c}_{e}}$)\footnote{This expression was derived by NTP's designers through years of experimentation and experience~\cite{nptfaq}.} before deciding to increase or decrease the polling interval (steps 6--14). This algorithmic detail implies that when we observe the same polling value for fewer than $N$ samples, we infer that the clock is going to a bad state ({\em i.e.}, losing synchronization) and assign the corresponding measurements to tier 2 (step 6 and 7). When we observe exactly $N$ samples in our logs with the same polling value followed by samples with an increased polling interval ({\em i.e.}, polling rate decreases), we infer that the $N$ samples must have been accepted by NTP's algorithm as good clock values and we therefore accept the $N$ samples too (steps 9 and 10). Similarly, $N$ samples followed by a decreased polling interval corresponds to clock values that we infer to be of poorer quality, thus we assign these $N$ samples to tier 2 (steps 11 and 12). If the number of samples is greater than $N$, we infer the client's clock to be in an unstable state, either shifting from a bad state to a good state (if polling interval increases), or from a good state to a bad state (if polling interval decreases). In either case, we cannot determine which samples are good and hence we assign all these samples to tier 2 (steps 13--14).

{\bf Constant polling behavior.} When a client sends a request to an NTP server, it sets the origin timestamp ($t_0$) to be equal to the transmit timestamp ($t_2$) from the previous server response.
%% from the NTP server. 
We refer to this behavior as {\em timestamp rotation}. Since our logs are captured at the server, we can obtain the server-to-client (s2c) latencies because of timestamp rotation. Similar to clients, servers also rotate timestamps when they send out an NTP response. Hence we can also estimate client-to-server (c2s) latencies. Timestamp rotation is an expected NTP protocol behavior and is used to prevent replay attacks (see~\cite{rfc5905}, p.28).

We can also recover the client-computed gtRTT between a client and a server which is reported by the client's {\tt ntpd} after correcting the system clock.  After the initial handshake between a client and a server, the client sets both the {\tt root delay} and {\tt ref id} fields in outgoing NTP request packets to the server's IP address and gtRTT estimate respectively. Thus, when we see a value in {\tt ref id} set to the IP addresses of one of our NTP log collection servers, we can get the client's estimate of gtRTT to our servers from the {\tt root delay} field. This offers an opportunity to enhance the filtering process. Furthermore, we found that for some client implementations the {\tt ref id} field is set to a wide variety of different NTP server addresses, thus providing client-computed RTT values between the client and multiple other servers.

We apply our observations on timestamp rotation behavior and the inclusion of gtRTT to filtering latency samples for clients that exhibit a constant polling interval.  First, we check if the packets between clients and servers contain gtRTT values (step 16).  Next, if the gtRTT values are present, we extract them from the root delay field and select only those packets in which the sum of OWDs is less than or equal to gtRTT (steps 17--20).  If the gtRTT value is absent, we default to the mean plus one-sigma deviation filter similar to prior work~\cite{Durairajan2015HotNets} (steps 21 and 22).  

%%%% \vspace{-0.2cm}
\begin{algorithm}[htb!]
\SetKwInOut{Input}{input}
\newcommand{\cmtsty}[1]{\texttt{\small #1}}\SetCommentSty{cmtsty}
$uList$ = []; \\
\BlankLine
\ForEach{client $C$ synchronizing with server $S$}{
	$type$ = classify($C$);\\
	\If{$type == non-constant$}{
		$N$ = ceil(30/$P^{c}_{e}$);\\
		\If{$nSamples < N$}{
			\tcp{clock going to bad state}
			%ignoreAndContinue;
			assignToTier2();
		}
		\ElseIf{$nSamples == N$}{
			\tcp{Extract N measurements}
			\If{$P^{n}_{e} > P^{c}_{e}$}{
				uList.append(takeNLatencies());\\
			}
			\ElseIf{$P^{n}_{e} < P^{c}_{e}$}{
				\tcp{clock going to bad state}
				%ignoreAndContinue;
				assignToTier2();
			}	
		}
		\Else{
			\tcp{clock is oscillating}
			%ignoreAndContinue;
			assignToTier2();
		}
	}
	\If{$type == constant$}{
		$flag$ = checkRootDelay();\\
		\If{$flag$}{	
			$gtRTT$ = getRootDelay();\\
			\If{$c2s + s2c <= gtRTT$}{
				uList.append(selectPackets());\\
			}
		}
		\Else{
			applyMeanFilter();\\ 
		}
	}
	\If{len($uList$) > 0}{	
		applyEWMA(uList);\\
		assignToTier3();\\
	}
	\If{one-shot requests}{	
		\If{$nSamples > 1$}{	
			\tcp{with OWD}
			assignToTier1();\\
		}
		\Else{
			\tcp{no OWD}
			assignToTier0();\\
		}
	}
}
\caption{NTP precision assignment algorithm}
\label{alg:filtering}
\end{algorithm}
%%%% \vspace{-0.2cm}

{\bf Sample smoothing.} Finally, even with these NTP-specific techniques for filtering OWD samples, there may yet be spikes that cause inaccuracies.
%% in the inferred latencies.
%% ({\em e.g.,} the two spikes in Figure~\ref{fig:filtering} between times 40000 and 50000). 
We apply a simple EWMA filter to smooth spikes in latency measurements in all tier 3 accepted samples. To choose the weighting factor ($\alpha$), we iterate from 0.1 to 0.9 such that the mean-squared error of filtered latencies is minimized (step 24).

Apart from the measurements with constant or non-constant polling value, a number of samples are either from SNTP clients with one-shot requests or from NTP clients sending (single or multiple) one-shot requests in our logs (steps 26 to 30). Since some or all of the timestamps are empty in the one-shot NTP and SNTP measurements, we assign them to tier 0 if OWD cannot be inferred (step 30). In addition, we found that a sizable number of measurements also exhibited similar behavior despite multiple one-way requests sent by NTP clients but with OWD information that we cannot verify. We assign such measurements to tier 1 (step 28).

{\bf Putting it all together.} We implemented our precision framework in about 1,250 lines of C++ code and applied it to our NTP traces. The result is a set of measurements assigned to one of the following four tiers. Details of the characteristics of OWDs in different tiers are described in \S\ref{sec:characteristics}, and the raw number of measurements assigned to each tier are depicted in Table~\ref{tab:statsTier}.

\squishlist
\item First, we assign all the one-shot measurements ({\em e.g.} SNTP) to {\em tier 0}. These include measurements with empty values in the timestamp field, partially filled timestamp fields, etc. We note again that we cannot infer any latency information from these measurements. 
\item Next, we assign the measurements that we filter using our heuristics-based technique to {\em tier 3}. OWD measurements in this tier are from well-synchronized clients. As a by-product of the tight synchronization between clients and references, we have a set of accurate client-to-server/server-to-client OWD measurements in this tier.
\item Next, those measurements that are rejected by our filter are assigned to {\em tier 2}. The measurements in this tier are from clients trying to achieve synchronization with references, but are not well-synchronized.  We note that we also apply a constant bound for measurements: if the OWDs are less than 1000ms, they are assigned into tier 2.
\item Finally, in the above tier 2 classification, OWDs greater than 1000ms are classified as {\em tier 1}. Furthermore, since we cannot infer the level of synchronization for a number of SNTP clients despite having multiple OWD samples, we include those measurements in {\em tier 1}.
\squishend

\begin{figure}[ht!]
  \centering
  \epsfig{figure=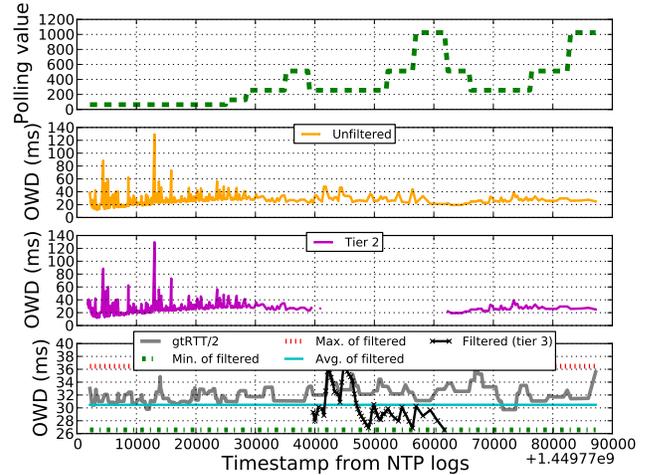,width=8.5cm}
  \caption{\label{fig:filtering}{\bf Latencies extracted from NTP logs from a lab-based client before and after applying filtering. Tier 3 and tier 2 measurements are shown in black and magenta, respectively.}}
\end{figure}

To illustrate concretely, Figure~\ref{fig:filtering} shows the {\em raw} polling values (top plot; dashed green curve) and {\em unfiltered} latencies (2nd to top plot; solid yellow curve) extracted from NTP logs for a client in our lab.  In addition to the polling values and unfiltered latencies, the figure shows gtRTT/2 (from the root delay field) extracted from NTP logs (bottom plot; grey curve).  Observe that between times 25000--37500 the polling interval increases and, indeed, there is a stabilization of OWDs to accurate values.  Likewise, between times 60000--70000 the polling interval decreases and there is some corresponding loss of stability in OWD samples.  The bottom two plots of Figure~\ref{fig:filtering} show that during time 40000--60000 and the stabilization of OWD (and likewise, the longest increasing trend in polling value), latency samples are classified and filtered as tier 3 measurements (steps 23 to 25) (bottom plot; black curve), while before and after that stable period samples are classified as tier 2 (3rd plot from top; magenta curve).  Also shown in the bottom plot are the min., max. and avg. of the tier 3 filtered latencies.  For this experiment, only tier 2 and 3 latencies resulted from processing the raw samples.
%% Between those regimes (between time 40000--60000, {\em i.e.}, the longest increasing trend in polling value), TimeWeaver selects and filters latency samples for tier 3 (bottom plot; black curve).  
%% The bottom plot of Figure~\ref{fig:filtering} also shows the tier 3 filtered latencies computed using Algorithm~\ref{alg:filtering} (in black; see from 40000 to 60000 {\em i.e.}, the longest increasing trend in polling value) and .  
%% While assigning measurements to tier 2, we also check if the OWDs are greater than 1s. Measurements with OWDs greater than 1s are assigned to tier 1; otherwise they are assigned to tier 2.

The tier 3 filtered samples shown in the bottom plot of Figure~\ref{fig:filtering} represent the most accurate estimates from the TimeWeaver framework.  We observe that the {\tt ntpd} client's computed latency value (gtRTT/2) aligns well with our filtered OWD estimate.  Notice also the figure 
%% Figure~\ref{fig:filtering} 
shows several spikes in unfiltered OWD samples ({\em e.g.}, two spikes between times 40000 and 50000) among other deficiencies, which are effectively addressed through the TimeWeaver framework.  Through extensive examination of many individual client traces, results for which are not shown here due to space constraints, we found that our filtering technique correctly and consistently eliminates poor samples and spikes that would otherwise pollute OWD estimates.

%%%% \vspace{-0.35cm}
%%%% \vspace{-0.35cm}

\begin{table*}[]
\centering
\caption{{\bf Number of measurements assigned to each tier by TimeWeaver.}}
\label{tab:statsTier}
\scalebox{0.8}{
\begin{tabular}{|l|l|l|l|l|l|l|l|l|l|l|l|l|l|l|l|l|l|l|l|} \hline
       & AG1 & CI1 & CI2 & CI3 & CI4 & EN1 & EN2 & JW1 & JW2 & MW1 & MW2 & MW3 & MW4 & MI1 & PP1 & SU1 & UI1 & UI2 & UI3 \\ \hline \hline
Tier 0 &1.5e8 & 9.4e6 & 9.7e6 & 1.1e7 & 1.1e7 & 5.4e6 & 5.5e6 & 3.0e6 & 1.7e7 & 2.4e6 & 5.2e8 & 1.8e8 & 2.6e8 & 5.5e8 & 4.7e6 & 2.4e8 & 1.9e8 & 6.7e7 & 2.7e7 \\
Tier 1 &1.3e8 & 5.2e6 & 5.2e6 & 4.9e6 & 5.4e6 & 4.9e5 & 8.5e5 & 2.8e6 & 1.6e7 & 1.1e6 & 2.6e8 & 1.1e8 & 1.1e8 & 4.6e8 & 4.0e6 & 2.2e8 & 3.6e7 & 5.2e7 & 2.0e7 \\
Tier 2 & 4.2e7 & 7.0e6 & 6.5e6 & 7.7e6 & 6.9e6 & 6.5e6 & 6.2e6 & 3.3e6 & 4.5e6 & 9.2e5 & 3.7e7 & 5.4e7 & 1.7e7 & 3.9e8 & 6.4e5 & 6.8e7 & 5.5e7 & 1.2e8 & 9.6e7 \\
Tier 3 & 3.0e7 & 1.5e6 & 8.4e5 & 1.4e6 & 7.0e5 & 7.8e5 & 8.4e5 & 2.4e6 & 2.9e6 & 1.0e6 & 1.3e7 & 1.7e7 & 6.0e6 & 5.1e8 & 7.8e5 & 5.6e7 & 1.9e7 & 3.6e7 & 3.2e7 \\ \hline
\end{tabular}
}
\end{table*}

%%%% \vspace{-0.2cm}
\subsection{Comparison with ping measurements}
%%%% \vspace{-0.2cm}

To assess the effectiveness of our approach, the administrators of several of the NTP servers (MW1-4) used {\tt ping} to send 10 probes each to a random sample of more than 20,000 client hosts identified in their logs. Ping measurements were issued simultaneously on a day that the NTP data was collected. We do not argue that ping measurements provide ground truth, rather that they provide a useful perspective on the NTP measurements. We include clients for which our algorithm assigned tiers 1, 2 or 3 for OWD samples. 6,370 out of 21,443 target clients responded to the pings.

%%%% \vspace{-0.3cm}
\begin{figure}[htb!]
  \centering
  \epsfig{figure=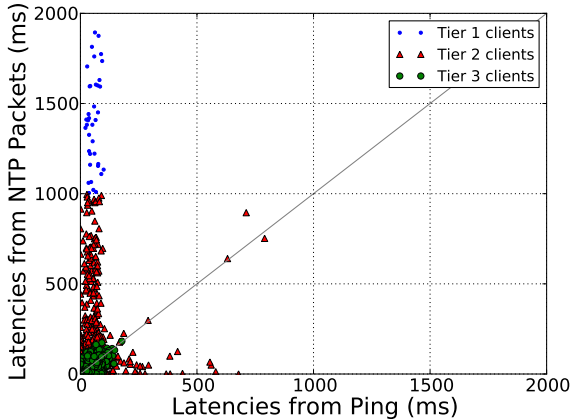,width=8cm}
  \caption{\label{fig:filteringValidation}{\bf Comparison of measured minimum $c2s + s2c$ latencies from NTP packets and RTT from ping measurements. Clients assigned to tier 3, 2 and 1 are denoted by green circles, red triangles, and blue dots, respectively.}}
\end{figure}
%%%% \vspace{-0.3cm}

Figure~\ref{fig:filteringValidation} shows a scatterplot comparison of the minimum of $s2c+c2s$ filtered latencies derived from NTP logs for the 6,370 clients compared with the corresponding minimum RTT values from ping measurements. Latencies from 3,708 clients were assigned to tier 3 by our framework and are shown as green circles, and 2,662 latency samples were in tiers 2 and 1 and are shown in red triangles and blue dots respectively. From these data points, we observe that there are no extreme outliers that are colored green. This indicates that our precision assignment approach is effective in assigning poor latency samples to lower tiers. Furthermore, a number of clients were assigned to lower tiers by Algorithm~\ref{alg:filtering} even though the $c2s+s2c$ latencies were comparable with the RTT measurements. On detailed examination of such clients, we found that the polling values were oscillating at the time when the packets were captured at these four servers. In such cases, without further information we must treat the latency samples as indeterminate and thus assign them to a lower tier. Overall, our results show that latency samples from NTP packet traces can indeed be used to derive OWD estimates of different precisions that are suitable for various applications.

%%%% \vspace{-0.2cm}
\subsection{TimeWeaver vs. prior NTP filtering} 
%%%% \vspace{-0.2cm}

A natural question is whether the filtering approach described in prior work (see \S4 in~\cite{Durairajan2015HotNets}), can be used to remove bad OWD measurements? To answer this question, we used the code from~\cite{Durairajan2015HotNets} and compared its filtering output vs. TimeWeaver.  Specifically, we randomly selected logs from our data corpus from multiple NTP servers across multiple days and compared the client and latency characteristics of TimeWeaver versus those produced from the prior method. Based on comparisons using one day's-worth of data from the JW1 server\footnote{Results from other days and other NTP servers exhibited similar characteristics.}, we found that the filtering approach used in~\cite{Durairajan2015HotNets} is not widely applicable for the following reasons:

{\bf Client characteristics.} {\em (1)} Out of the 18,620 unique clients seen in the log of JW1 server on a randomly-selected day,~\cite{Durairajan2015HotNets} only considers 8,804 clients due to its US-only filtering constraint. On the contrary, TimeWeaver 
%% goes well beyond~\cite{Durairajan2015HotNets} and 
considers all 18,620 clients spread across many countries (see \S\ref{sec:characteristics} for client characteristics of the entire JW1 dataset). {\em (2)} Of the 8,804 clients considered by the prior method, a large fraction of clients ({\em i.e.}, about 3,631) were rejected due to missing timestamps, negative latency values, and other reasons. TimeWeaver, on the other hand, assigns such discarded measurements to lower tiers, making it possible to use less accurate OWD values in applications that have less stringent accuracy requirements.
%% ({\em cf.} \S\ref{sec:application}).
%% enhancing the utility of latency measurements derived from NTP 
%% overcoming the downsampling issue and enhances the utility of OWDs and clients in lower tiers (see \S\ref{sec:application}).

\begin{figure*}[!htb]
\begin{centering}
\epsfig{figure=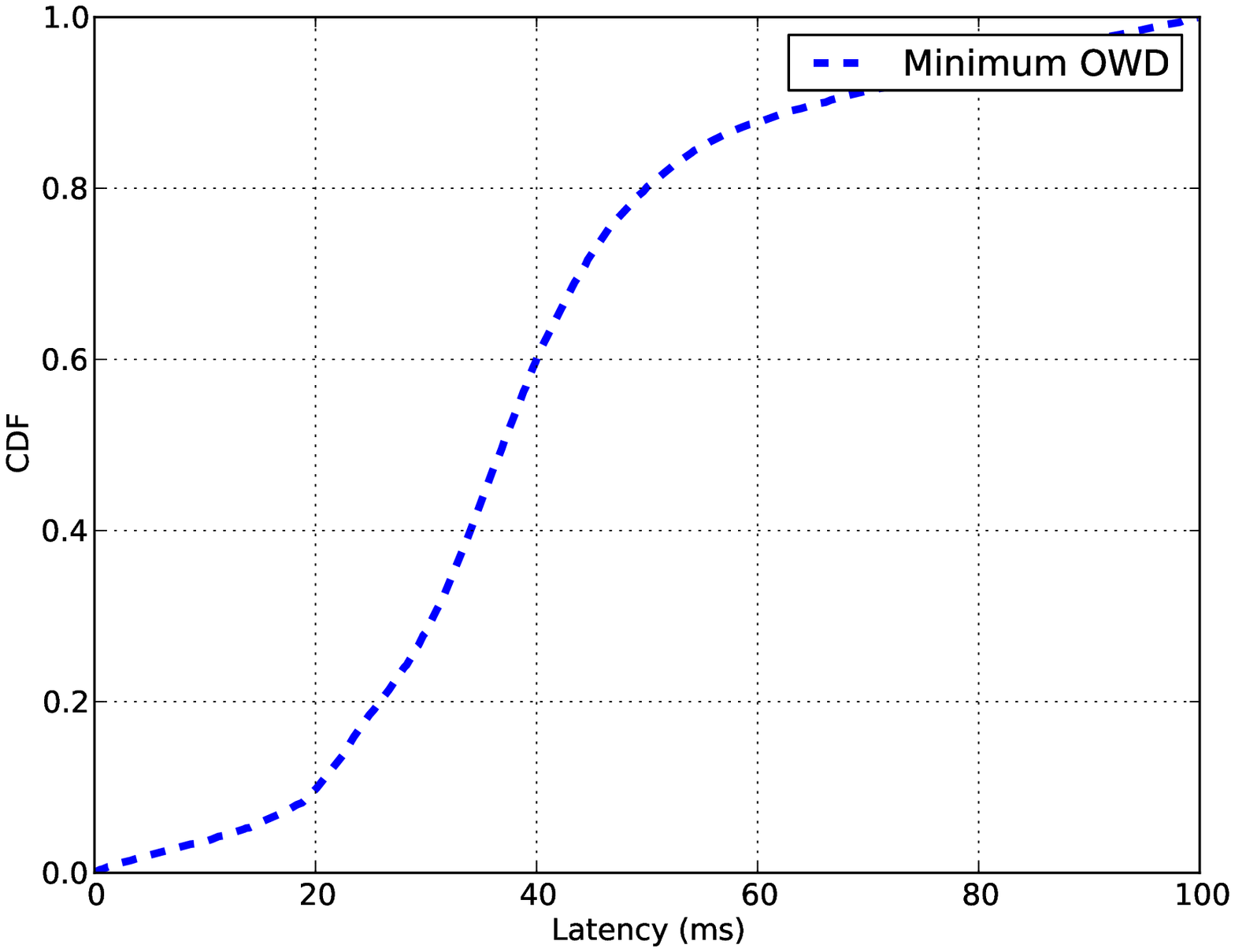,width=8cm}
\epsfig{figure=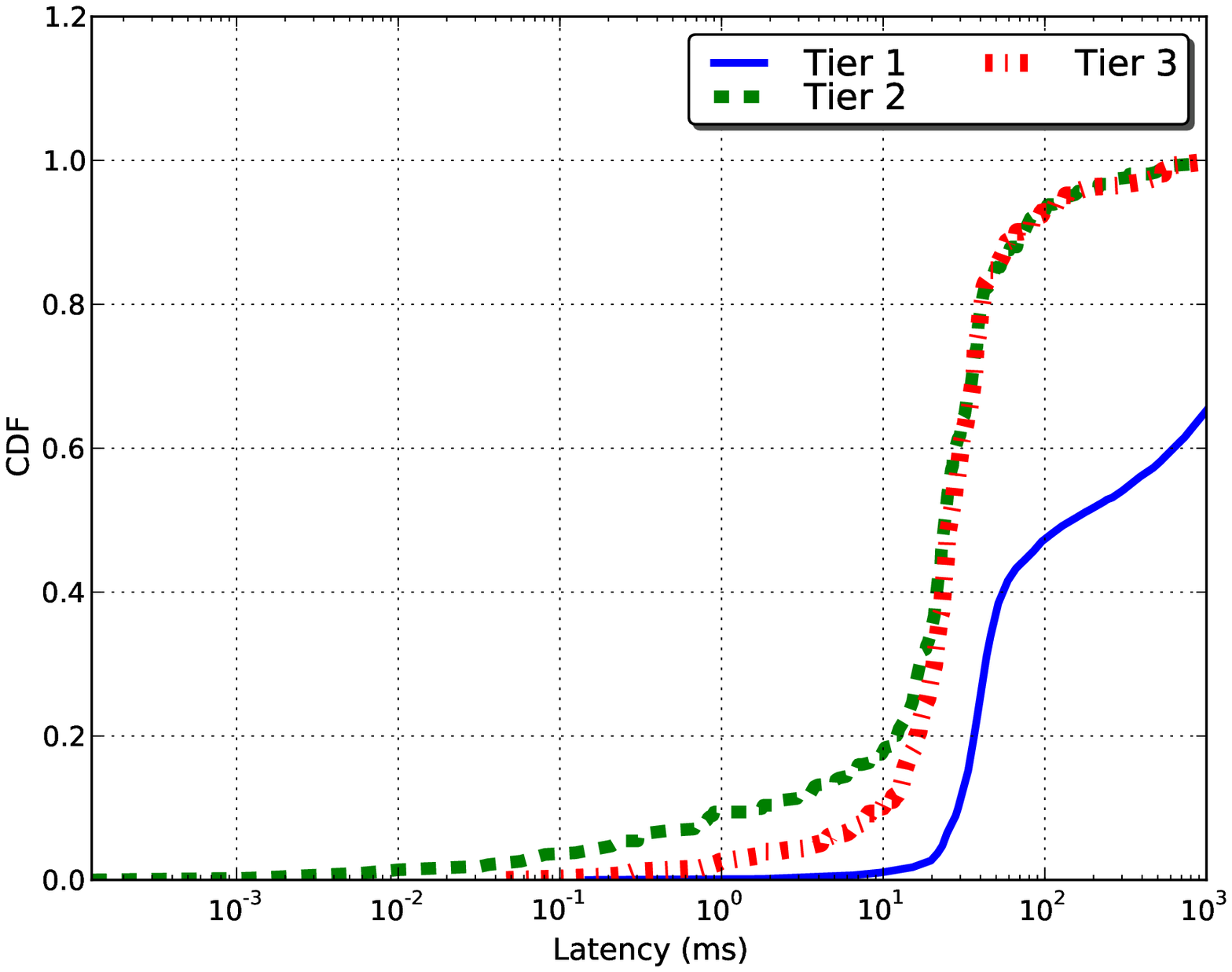,width=8cm}
  {\caption{\label{fig:latDistHotOrNot}{{\bf CDF of minimum OWD latencies extracted using Durairajan {\em et al.}~\cite{Durairajan2015HotNets} (left) versus TimeWeaver (right). Note that the x axis for TimeWeaver (right) is cut at 1000ms to make the plots more comparable.}}}}
\end{centering}
\end{figure*}

{\bf Latency characteristics.} Apart from considering the US-only clients,the method from~\cite{Durairajan2015HotNets} also uses a 100ms OWD threshold to limit wired vs. wireless hosts. As a consequence, 
%% of this constraint, 
the observed latency characteristics after applying TimeWeaver (right) are completely different from~\cite{Durairajan2015HotNets} (left), as depicted in Figure~\ref{fig:latDistHotOrNot}. First, the maximum of the extracted minimum OWD is 100ms in prior work, whereas the maximum value for a client in TimeWeaver's tier 3 category is 992ms.  Second, about 80\% of the clients filtered using~\cite{Durairajan2015HotNets} exhibited a latency less than 50ms, while only 14\% of the tier 3 clients had OWDs less than 50ms using TimeWeaver due to a more flexible and NTP-specific filtering approach.

%% file: characteristics.tex
%%%% \vspace{-0.2cm}
In this section, we provide an analysis of the general characteristics of one-way latency as revealed through our NTP log data and TimeWeaver's tiered filtering approach. %Space constraints limit our report to high-level characteristics.

{\bf Scope and reach of clients.}  We examine the geographic reach and locations of clients seen in NTP logs and observe that the client base for each server is, in general, large and widely distributed as seen in Table~\ref{tab:reach}. For example, for the SU1 server, clients make requests from 238 countries\footnote{We note that the raw counts vary across sources ({\em e.g.}, 194 countries in~\cite{source194}, 195 countries in~\cite{source195}, 196 countries in~\cite{source196}, 247 countries in~\cite{source247}). In this study, we report the raw counts based on MaxMind's free IP geolocation database which follows counting countries~\cite{maxMindSource} similar to~\cite{source247}.} across the world (identified via MaxMind). Similarly, a large number of servers ({\em e.g.}, AG1 and MI1), have clients spread across nearly all countries of the world. Interestingly, the MI1 stratum-1 server handles requests for a broadly distributed client pool, similar to high-traffic stratum-2 servers, {\em e.g.}, AG1 and SU1. We found this surprising because administrators of stratum-1 servers typically restrict the set of allowed clients.

%%%% \vspace{-0.3cm}
\begin{table}[!htbp]
\centering
  {\caption{\label{tab:reach}{{\bf Summary of scope and reach of clients in NTP server logs.}}}}
\begin{tabular}{ |c|c|c|c| }
\hline
Server ID	&	Countries	&	Regions/States		&	Cities 			\\ \hline \hline
AG1		&	238 		& 	391 				& 	49,978	  		\\ 
CI1		&	44 		& 	88 				&	279 	 			\\
CI2		&	47 		& 	86 				& 	260  				\\
CI3		&	22 		& 	61 				& 	196  				\\
CI4		&	24 		& 	45 				& 	134  				\\ 
EN1		& 	22 		& 	46 				& 	153  				\\
EN2		& 	18 		& 	44 				& 	132  				\\ 
JW1		& 	207 		& 	315 				& 	13,819	  		\\
JW2		& 	210 		& 	388 				& 	23,091  			\\ 
MW1		& 	134 		& 	351 				& 	6,156 	 		\\
MW2		& 	218 		& 	386 				& 	42,256	  		\\
MW3		& 	214		& 	384 				& 	37,573	  		\\
MW4		& 	222 		& 	389 				& 	39,226	  		\\ 
MI1		& 	237 		& 	388 				& 	39,686	  		\\ 
PP1		& 	218 		& 	361 				& 	19,347			\\ 
SU1		& 	238 		& 	389 				& 	52,444  			\\ 
UI1		&	156 		& 	347 				& 	9134		  		\\
UI2		&	164 		& 	353 				& 	11032	 		\\ 
UI3		&	148 		& 	343 				& 	8888  			\\ \hline
\end{tabular}
\end{table}
%%%% \vspace{-0.3cm}

\begin{figure}[htb!]
\begin{centering}
\epsfig{figure=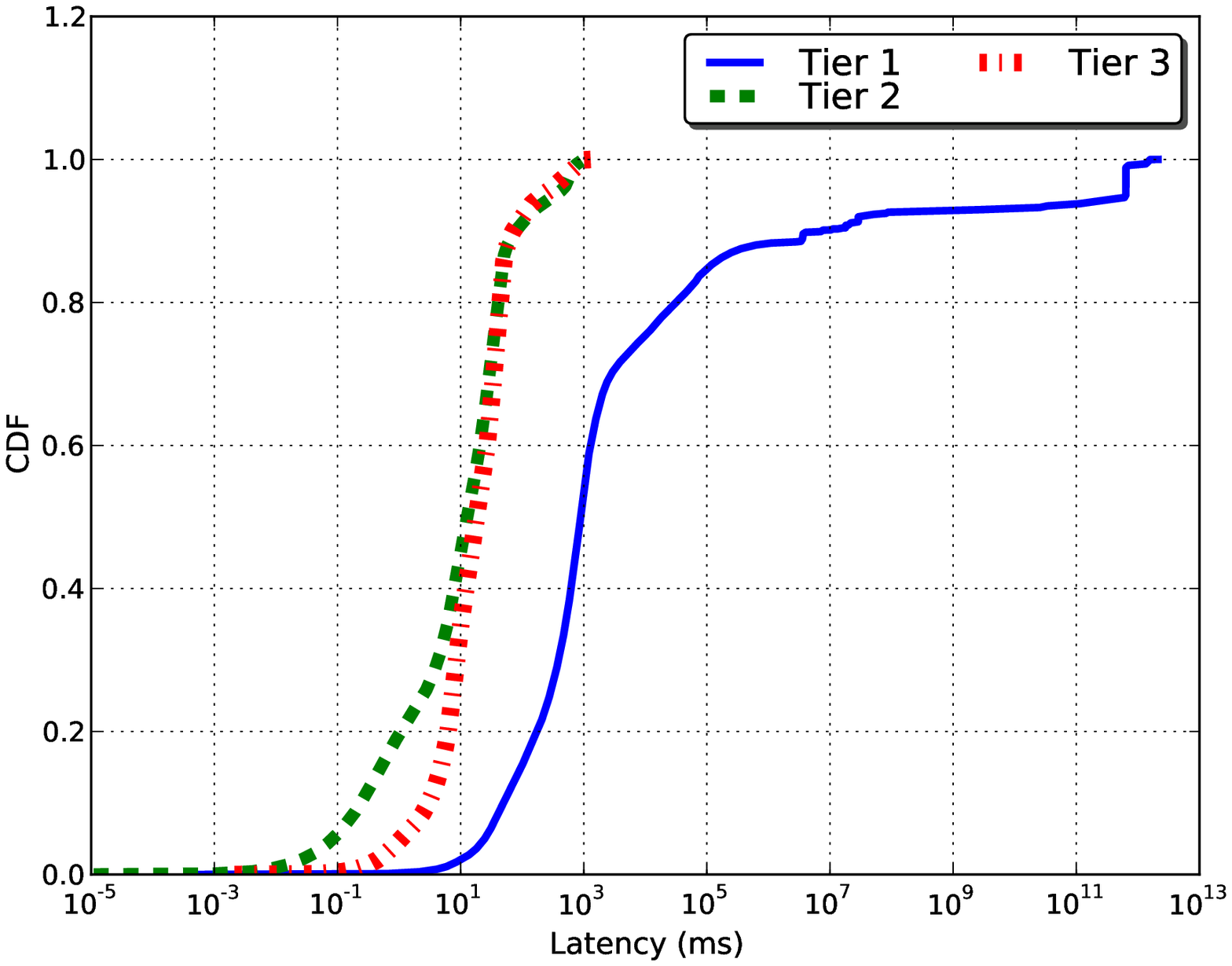,width=8.5cm}
\epsfig{figure=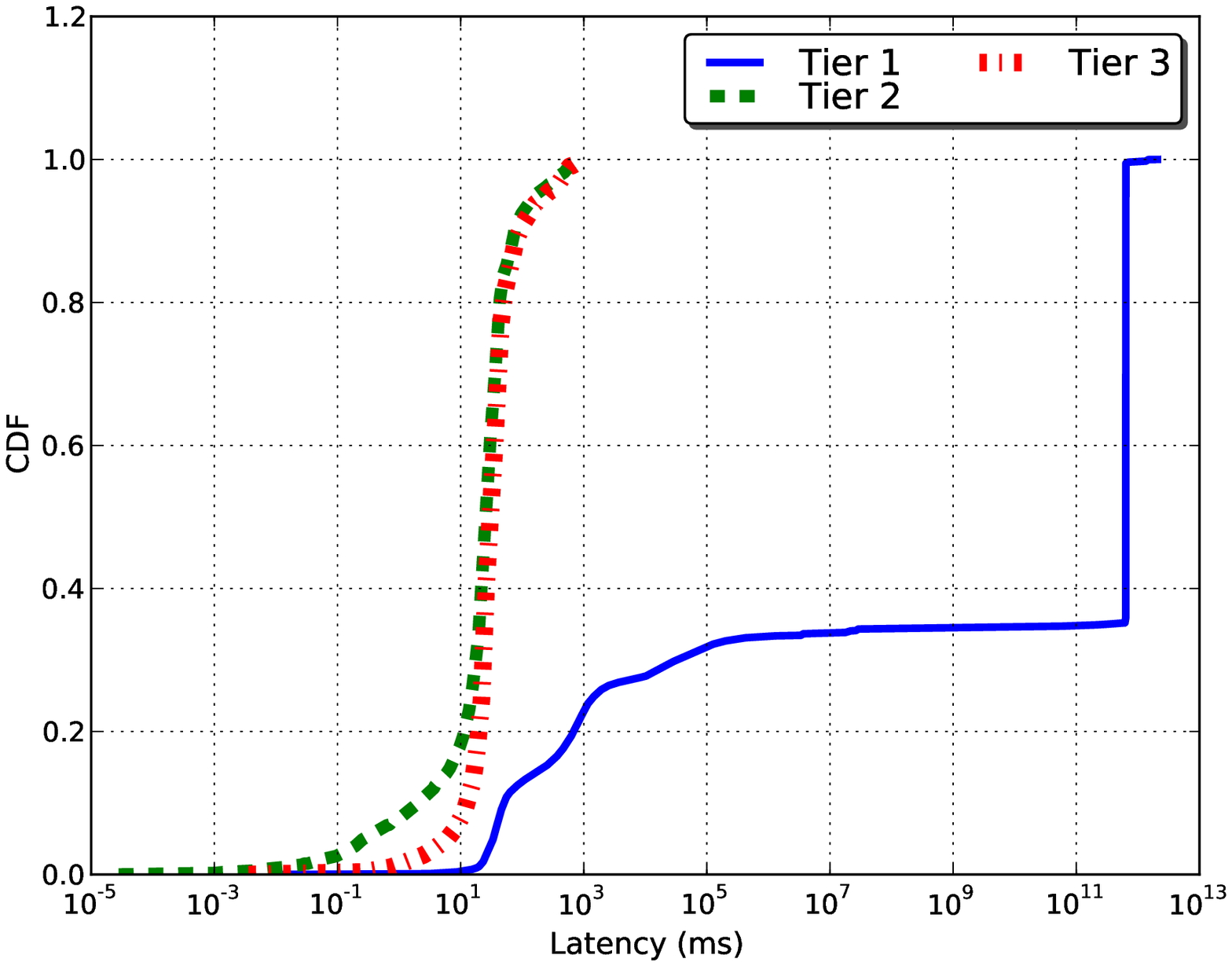,width=8.5cm}
\epsfig{figure=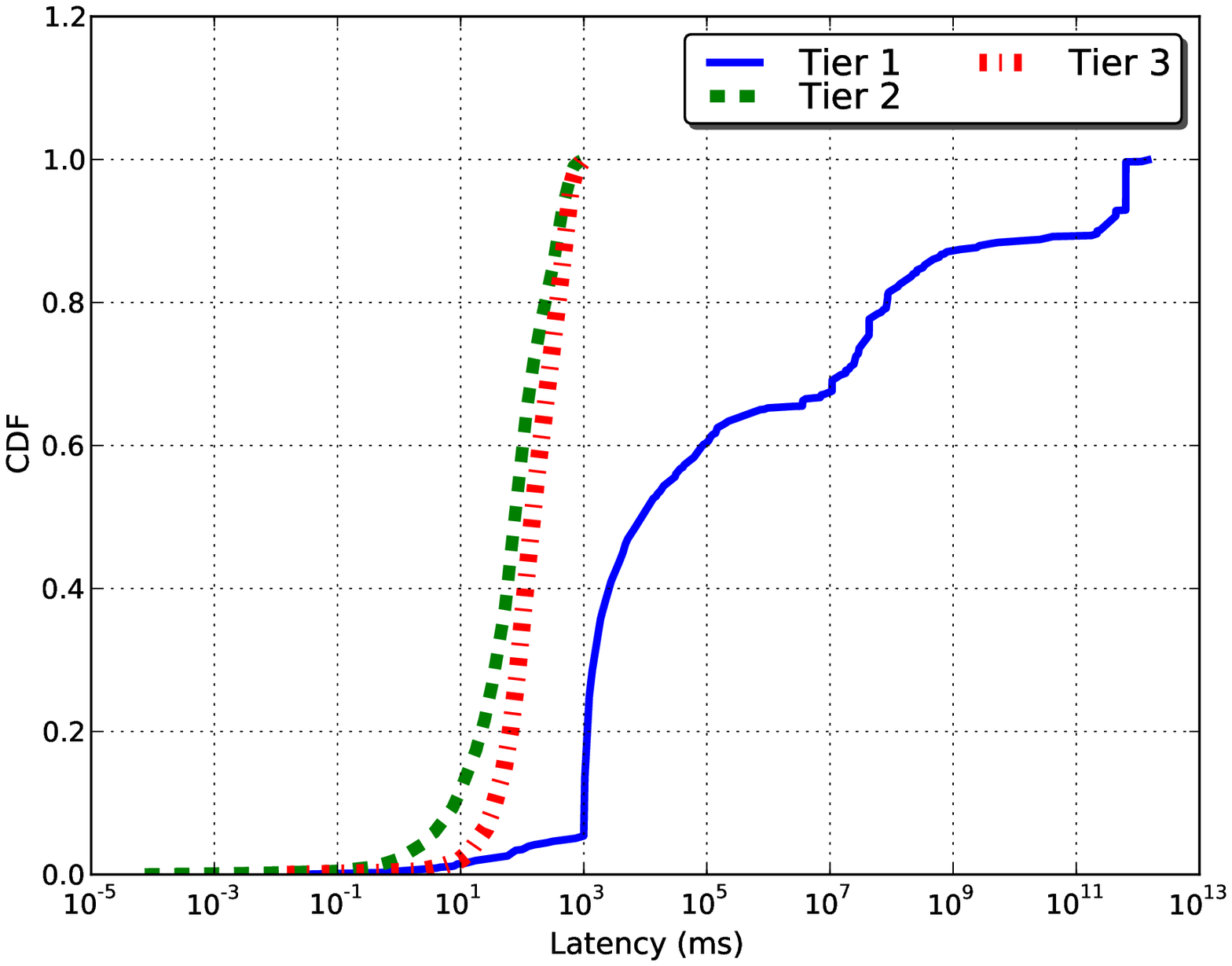,width=8.5cm}
  {\caption{\label{fig:latDist}{{\bf CDF of minimum OWD latencies for AG1 (top), JW2 (middle), and MW1 (bottom) NTP servers.}}}}
\end{centering}
\end{figure}

{\bf Latency distribution.} Figure~\ref{fig:latDist} shows the empirical CDF of minimum OWDs for a representative subset of servers ({\em i.e.,} AG1, JW2, and MW1). From these figures, we first observe that the TimeWeaver framework effectively assigns OWD measurements from out-of-sync clients to lower tiers. Specifically, OWDs as precise as (or below) 100ns are possible {\em only} with clients using Precision Time Protocol~\cite{ieeeptp} and are atypical of clients using NTP~\cite{millsPTP}. Thus the OWDs, as low as $10^{-5}$ms, can only be attributed to clients whose clock is leading with respect to their NTP reference, which are assigned to tier 2 category by TimeWeaver. 

From Figure~\ref{fig:latDist}, we also observe that 50\% of well-synchronized (tier 3) clients have OWDs less than 100ms. In contrast, only about 10\% of the clients in tier 1 category have OWDs less than 100ms, with many tier 1 clients exhibiting extremely large latencies (99th percentile is about $10^{12}$ms for each) indicative of poor synchronization.

%% file: matrix.tex
%%%% \vspace{-0.2cm}
In this section we introduce an algorithm for predicting unobserved latencies between clients, as well as predicting latencies between NTP servers and clients that were not measured.   The predictions are based on the tier 3 subset of OWDs extracted through TimeWeaver's precision assignment algorithm. 
%% measured between NTP servers and a subset of clients.  
A key observation we make is that a matrix of Euclidean distances between points in the $2$-dimensional plane has rank $4$.  The matrix of geodetic distances~\cite{geodetic} on the sphere is not exactly low-rank, but is well-approximated by a low-rank matrix.  This implies a significant level of correlation must exist among the pairwise latencies, which our algorithm exploits. 

%%%% \vspace{-0.2cm}
\subsection{Problem setup}
%%%% \vspace{-0.2cm}

We organize the latency matrix $X$ in the following block-form. Given $m$ NTP servers and $n$ clients, the latency matrix is $(m+n) \times (m+n)$ and can be arranged as follows
\[
X=
  \begin{bmatrix}
    A & B \\
    B^T & C
  \end{bmatrix}
\]
where $A$ is the $m$ x $m$ sub-matrix corresponding to inter-NTP server latency measurements, $B$ is the $n$ x $m$ block that corresponds to the c2s from $m$ NTP servers to $n$ clients, and $B^T$ is the $m$ x $n$ sub-matrix corresponding to the s2c distances from $n$ clients to $m$ NTP servers. Distance estimates from both $B$ and $B^T$ are extracted from the NTP logs and are partially incomplete sub-matrices.  The matrix $C$ corresponds to the c2c distances, and is completely unobserverd.

Although $C$ is completely unobserved, it may still be possible to estimate these latencies from the measured data.  To understand why, suppose that we fully observe $A$ and $B$. If $\mbox{rank}(X)=4$ (as discussed above), $m>=4$, and the first $m$ rows/columns have rank $4$, then we can complete $C$ with 

\vspace{-0.45cm}
\begin{equation} \label{completion_eqn}
	C = B A^{\dagger} B^T
\end{equation}
\vspace{-0.45cm}

where $A^{\dagger}$ is the {\em pseudo-inverse} of $A$. 

Unlike prior efforts on distance estimation that assume a full matrix of RTT measurements~\cite{dabek2004, mao2006ides, chen2009phoenix, liao2010network}, the situation we face is more challenging. The matrix $B$ is only partially observed in the logs and the matrix $A$ is not observed at all. To deal with this, we propose the following: {\em (1)} estimate the latencies in $A$ from the {\em known} physical distances between NTP servers, and {\em (2)} employ a low-rank matrix completion method to deal with the missing entries in $B$ and $C$.

%\vspace{-0.3cm}
\subsection{Estimating inter-NTP server distances}
%%%% \vspace{-0.1cm}

To estimate the distances between the $m$ NTP servers, we use the following approach.  First, from the NTP pool server list~\cite{ntppool}, we obtain the physical coordinates of the $m$ NTP servers used in our study.  Using the Vincenty formula~\cite{vincenty}, we compute the line-of-sight physical distances between all NTP servers. We calculate the speed-of-light estimate of latencies as roughly ${2/3}^{rd}$ the speed of light in air~\cite{singla2014} to obtain, $A_{geo}$, the geo-based estimate of $A$.

Next, we reached out to the NTP server administrators to measure the inter-NTP server latencies; 5 administrators (managing 11 servers) responded positively.  We performed 10 ping measurements from each of the 11 NTP servers and used ${rtt/2}$ of the minimum of ping measurements to create $A_{rtt}$, the rtt-based estimate of A. 

Lastly, since $A_{rtt}$ is partially incomplete and $A_{geo}$ only gives the {\em ideal} lower bound of distances, we use a scaling factor, $\gamma$, to obtain $A$.  To derive $\gamma$, we use a simple linear model to capture the sharing of information in the data ({\em e.g.}, all MW servers are located in Madison, WI) and $\gamma$ is based on linear model coefficients ${\beta}_{0}$ and ${\beta}_{1}$. That is, the coefficients are obtained by solving a simple linear regression ($y$ = ${\beta}_{0}$ + ${\beta}_{1}x$) for non-zero entries in $A_{rtt}$ and using the obtained ${\beta}_{0}$ and ${\beta}_{1}$ on $A_{geo}$ for those measurements for which we do not observe distances in $A_{rtt}$. The final inter-NTP distance matrix $A$ is a combination of $A_{rtt}$ and $\gamma$ applied on $A_{geo}$. 

%%%% \vspace{-0.3cm}
\subsection{Distance estimation algorithm}
%%%% \vspace{-0.1cm}

The distance estimation algorithm we employ is based on iterative hard-threshold singular value decomposition (IHTSVD)~\cite{chunikhina2014}. This is an iterative algorithm that alternates between truncating the SVD of the current estimate to a user-specified rank $k$, and then replacing the values in the observed entries with their original (observed) values. The algorithm can be initialized by setting the missing entries in $X$ arbitrarily.  In our experiments we initialize the missing values with the mean latency in the NTP logs.  Because Euclidean distance matrices in $2$-dimensions have rank $4$, we set $k=4$ in the algorithm. We apply the algorithm to tier 3 minOWD values from NTP clients that contact four or more servers. Since mobile and wireless clients can confound our estimation, we identify and remove those clients using Cymru lookup~\cite{cymru}, to the extent possible.

%%%% \vspace{-0.3cm}
\subsection{Assessing predicted distances}
%%%% \vspace{-0.3cm}

\begin{figure}[htb!]
 \centerline{\epsfig{figure=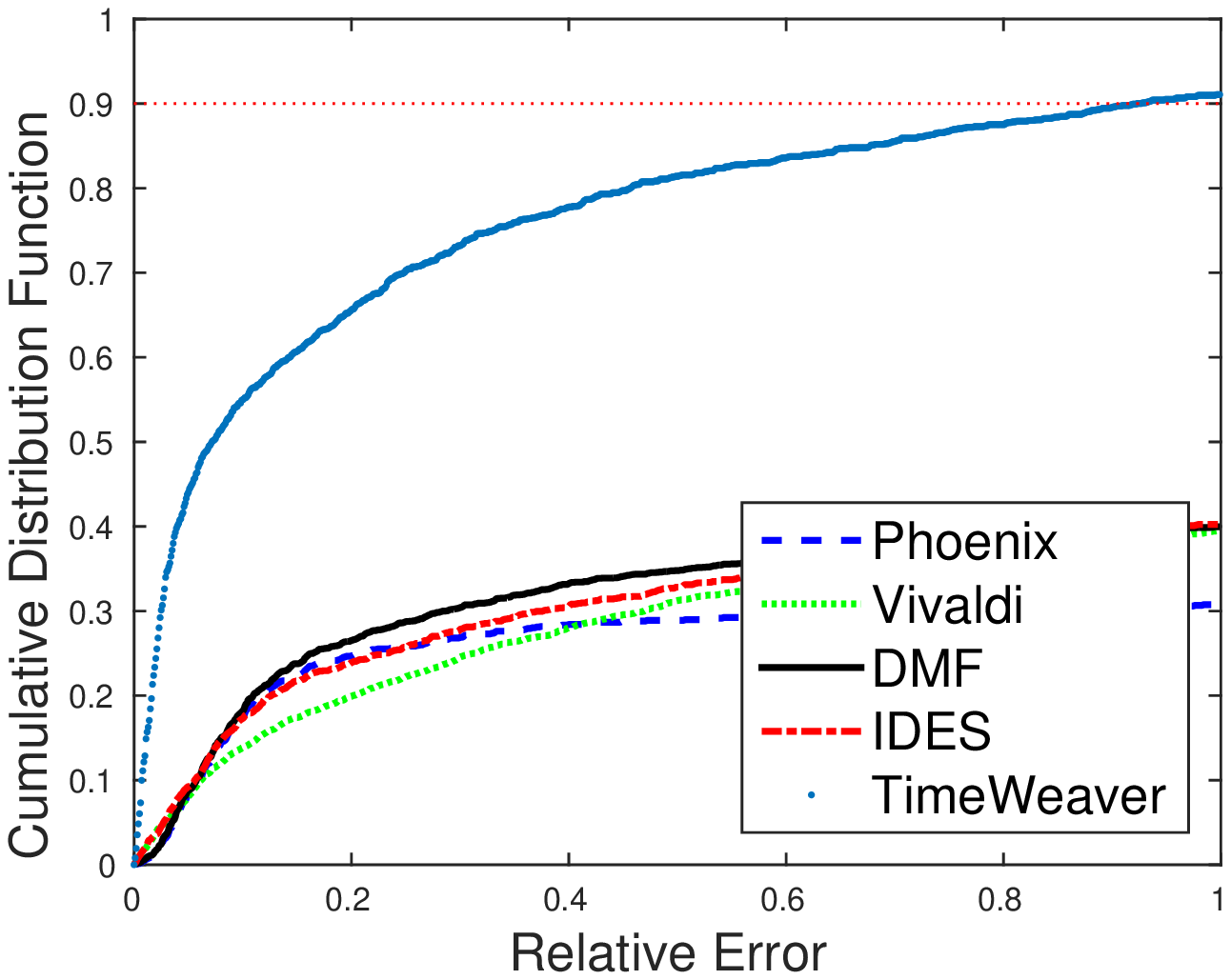,width=8.5cm}}
  \centerline{\epsfig{figure=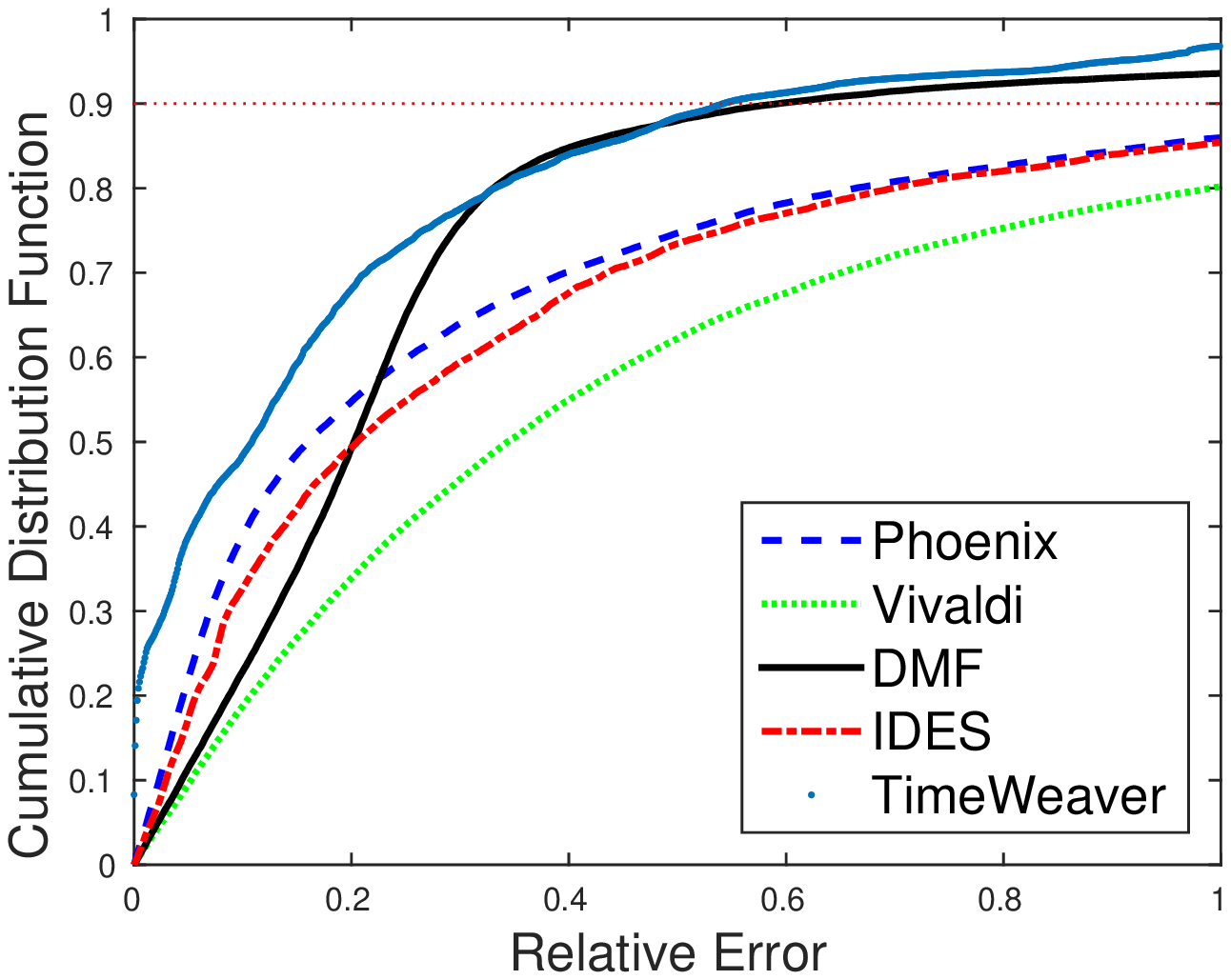,width=8.5cm}}
\caption{\label{fig:comparisonOthers}{{\bf CDF of relative errors using our approach (labeled TimeWeaver), Vivaldi, Phoenix, DMF and IDES after matrix completion for incomplete (top) and completed (bottom) data.}}}
\end{figure}

{\bf Comparison with other techniques.}  We compare the relative error of the our distance estimates against prior distance estimation techniques including Vivaldi~\cite{dabek2004}, IDES~\cite{mao2006ides}, Phoenix~\cite{chen2009phoenix}, and DMF~\cite{liao2010network}.  Figure~\ref{fig:comparisonOthers}-(top) shows the CDF of relative error made by TimeWeaver-based predictions versus the other methods for predicting the missing values in incomplete matrix $X$, where the sub-matrix $B$ is partially observed and sub-matrix $A$ is completely unobserved. The same OWD data is used in all cases.  For 50\% of the estimates, TimeWeaver-based predictions were off by at most 6\% from the original values, while for {\em only} 12\% of the estimates, similar relative errors were achievable using the other distance estimation techniques.

%% \begin{figure}[htb!]
%% \begin{centering}
%% \begin{minipage}[h]{0.49\linewidth}
%%  \centerline{\epsfig{figure=Figures/rerr_incomplete_data.eps,width=4.5cm}}
%% \end{minipage} 
%% \begin{minipage}[h]{0.49\linewidth}
%%   \centerline{\epsfig{figure=Figures/rerr_completed_data.eps,width=4.5cm}}
%% \end{minipage}  
%% \caption{\label{fig:comparisonOthers}{{\bf CDF of relative errors for TimeWeaver, Vivaldi, Phoenix, DMF and IDES after matrix completion for incomplete data (left) and completed data (right).}}}
%% \end{centering}
%% \end{figure}

%%%% \vspace{-0.2cm}
%%%% \vspace{-0.2cm}

Figure~\ref{fig:comparisonOthers}-(bottom) shows the CDF of relative error for predictions from the same set of entities (mentioned above) but for the completed matrix $X$. In this analysis, the estimates were randomly held-out and then predicted again. The plots show that TimeWeaver-derived distance estimates are perfectly accurate for 20\% of the estimates, and have a relative error of 10\% for 50\% of the estimates. For 80\% of the estimates, the predictions are off by 34\% and beyond that the results are comparable with DMF.  

We approach latency/distance estimation as a low-rank matrix completion problem. The basic ingredients in the algorithm (matrix factorization) are also used in the prior methods.  However, our approach has features that can explain its superior performance.
%% in the experiments. 
Vivaldi aims to explicitly determine a low-dimensional embedding of the network that agrees with the measured latencies; we target distance estimation directly.  In this sense, our approach is similar to DMF, although we use a centralized global optimization procedure and do not require regularization beyond that imparted by the low-rank constraint. IDES is also similar, but is landmark-based and assumes few if any missing measurements. In contrast, our approach is designed to handle cases in which most of the data are missing.
%As we observed previously, the tier 3 minOWD values derived through TimeWeaver are approximately line-of-sight, which we argue should minimize the likelihood of TIV---an issue that has arisen with prior efforts that have used RTT-based delay measurements.  
The results in this section show that even when prior methods use minOWD, predictions using our approach are more accurate, especially for the situation in which we have incomplete data.  

\begin{figure*}[htb!]
\begin{centering}
\begin{minipage}[h]{0.33\linewidth}
\centerline{\epsfig{figure=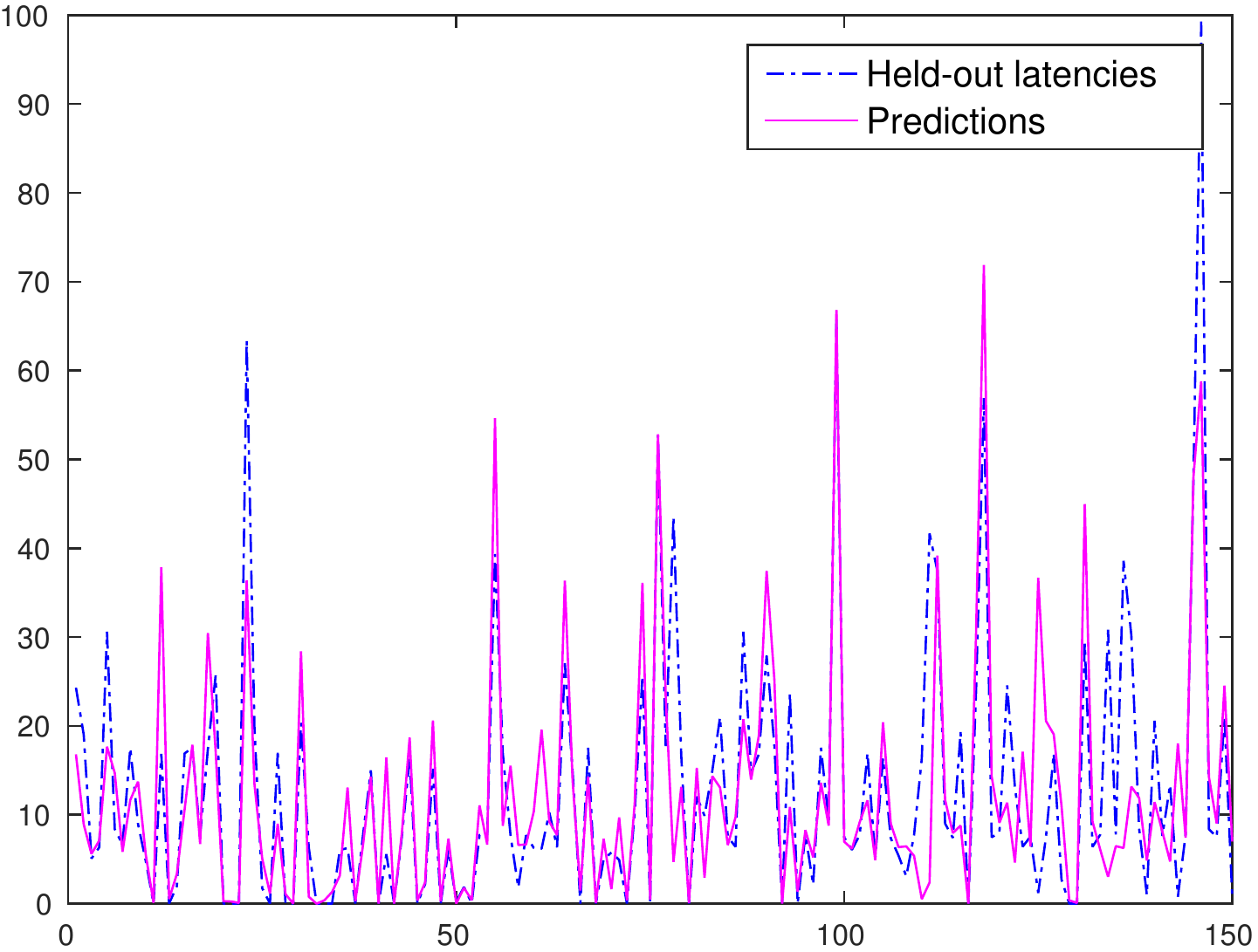,width=5.7cm}}
% \centerline{\epsfig{figure=Error_Plots/error_style2_100.eps,width=5.7cm}}
\end{minipage} 
\begin{minipage}[h]{0.33\linewidth}
\centerline{\epsfig{figure=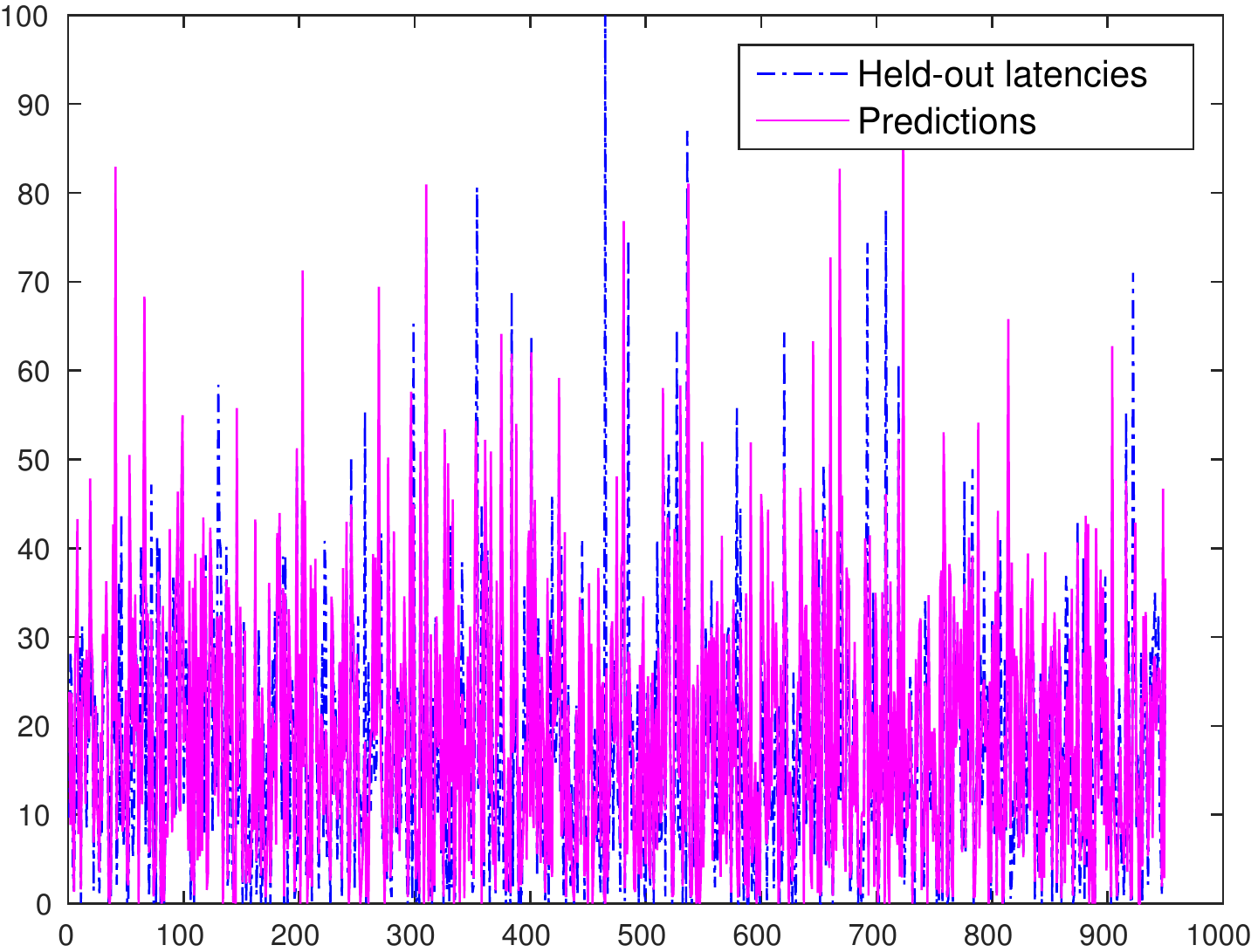,width=5.7cm}}
% \centerline{\epsfig{figure=Error_Plots/error_style2_1000.eps,width=5.7cm}}
\end{minipage}  
\begin{minipage}[h]{0.33\linewidth}
\centerline{\epsfig{figure=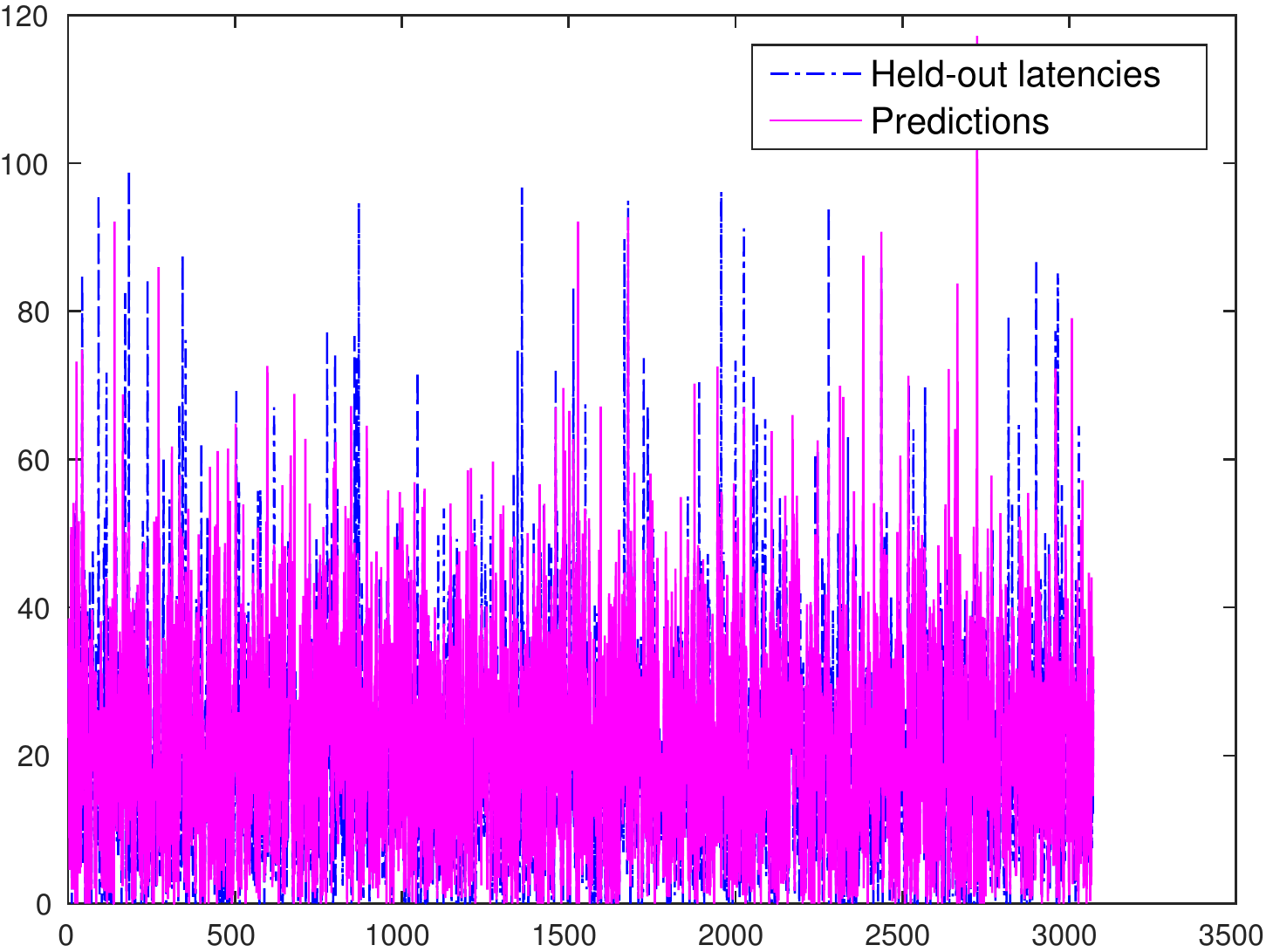,width=5.7cm}}
% \centerline{\epsfig{figure=Error_Plots/error_style2_4000.eps,width=5.7cm}}
\end{minipage} 
\caption{\label{fig:holdOut}{{\bf Average relative prediction error (hold-out distances vs. predicted distances) for 100 (left), 1000 (center) and 3668 (right) clients.}}}
\end{centering}
\end{figure*}

\begin{figure*}[htb!]
\begin{centering}
\begin{minipage}[h]{0.33\linewidth}
% \centerline{\epsfig{figure=Figures/error_100.eps,width=5.5cm}}
\centerline{\epsfig{figure=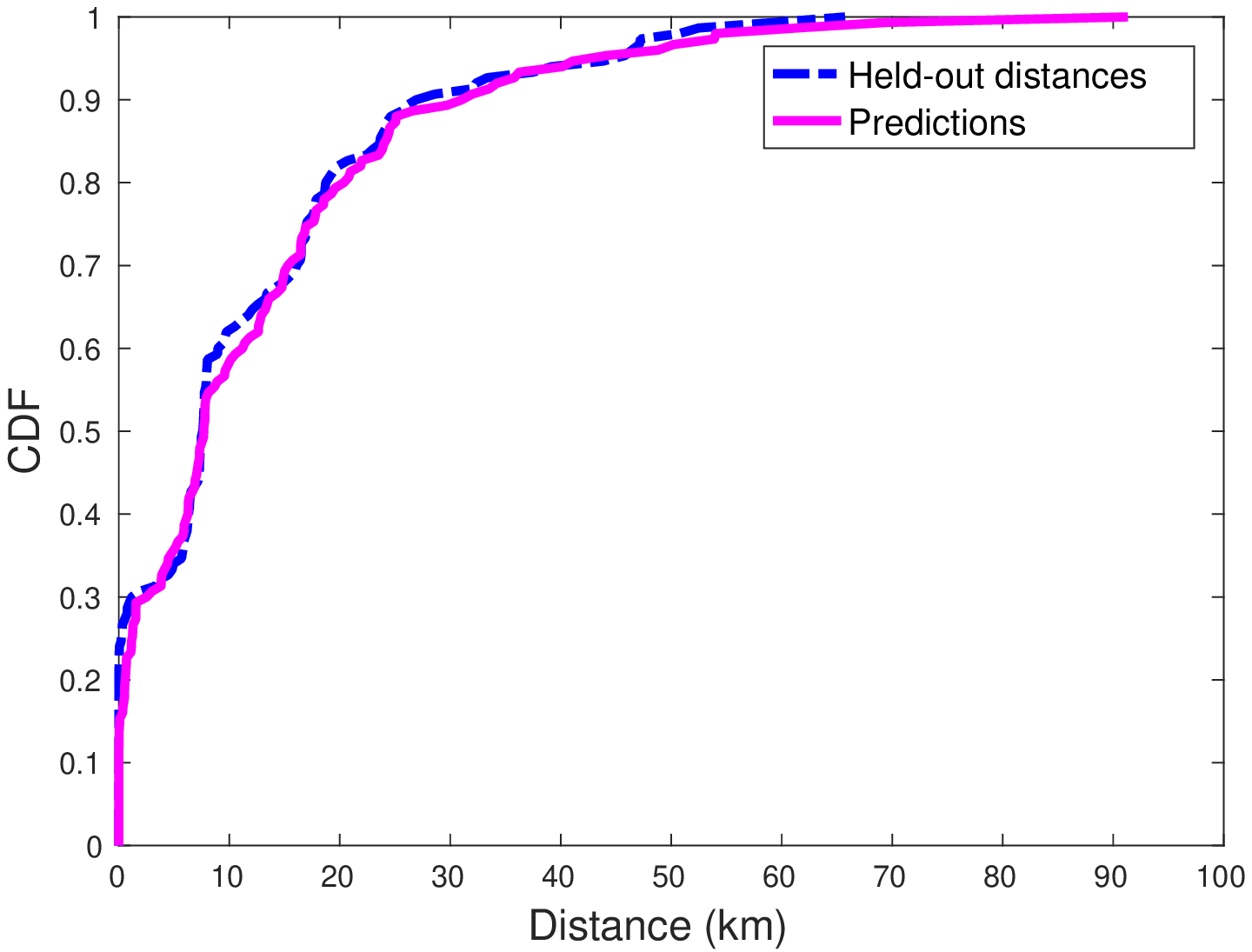,width=5.7cm}}
\end{minipage} 
\begin{minipage}[h]{0.33\linewidth}
%  \centerline{\epsfig{figure=Figures/error_1000.eps,width=5.5cm}}
\centerline{\epsfig{figure=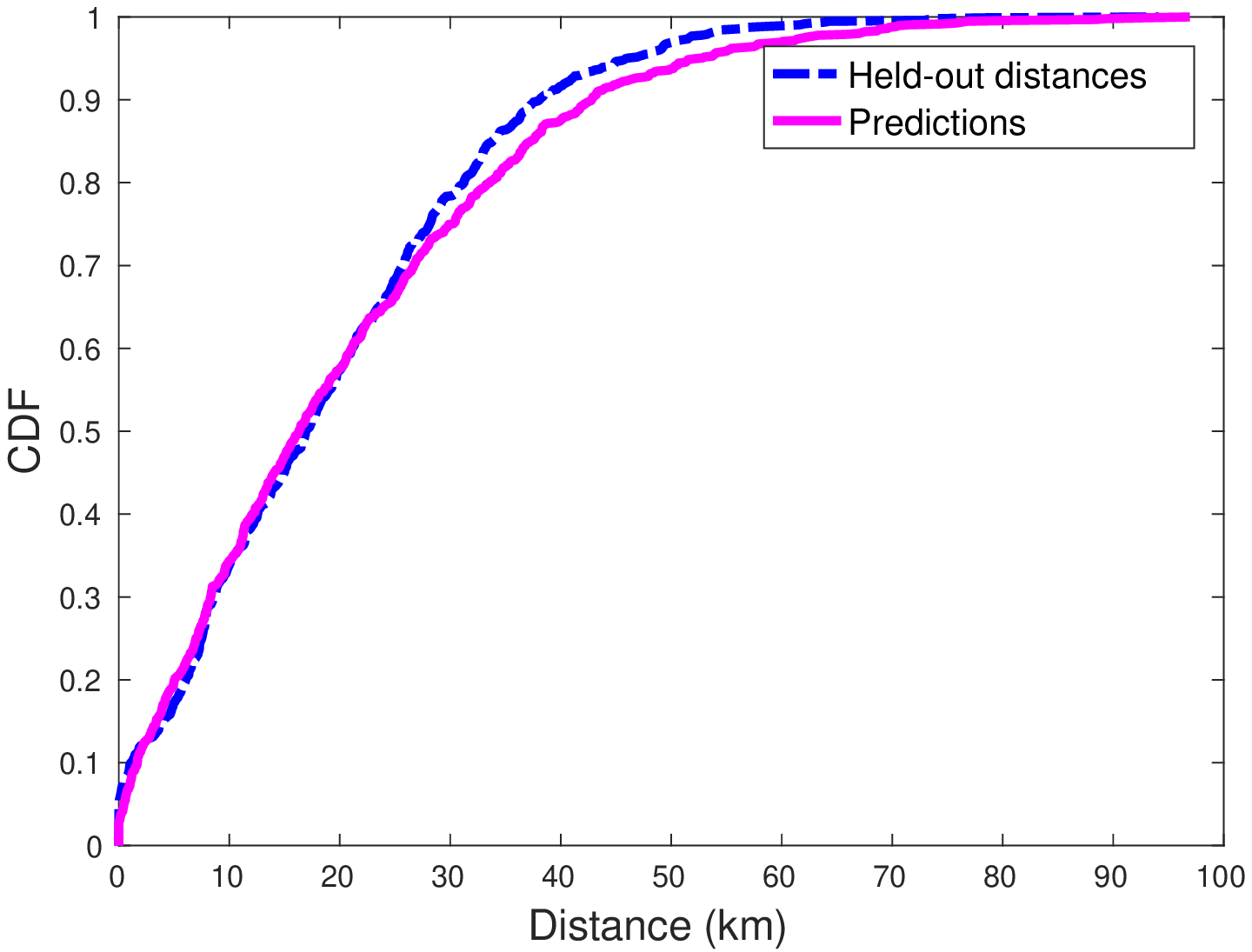,width=5.7cm}}
\end{minipage}  
\begin{minipage}[h]{0.33\linewidth}
%  \centerline{\epsfig{figure=Figures/error_4000.eps,width=5.5cm}}
\centerline{\epsfig{figure=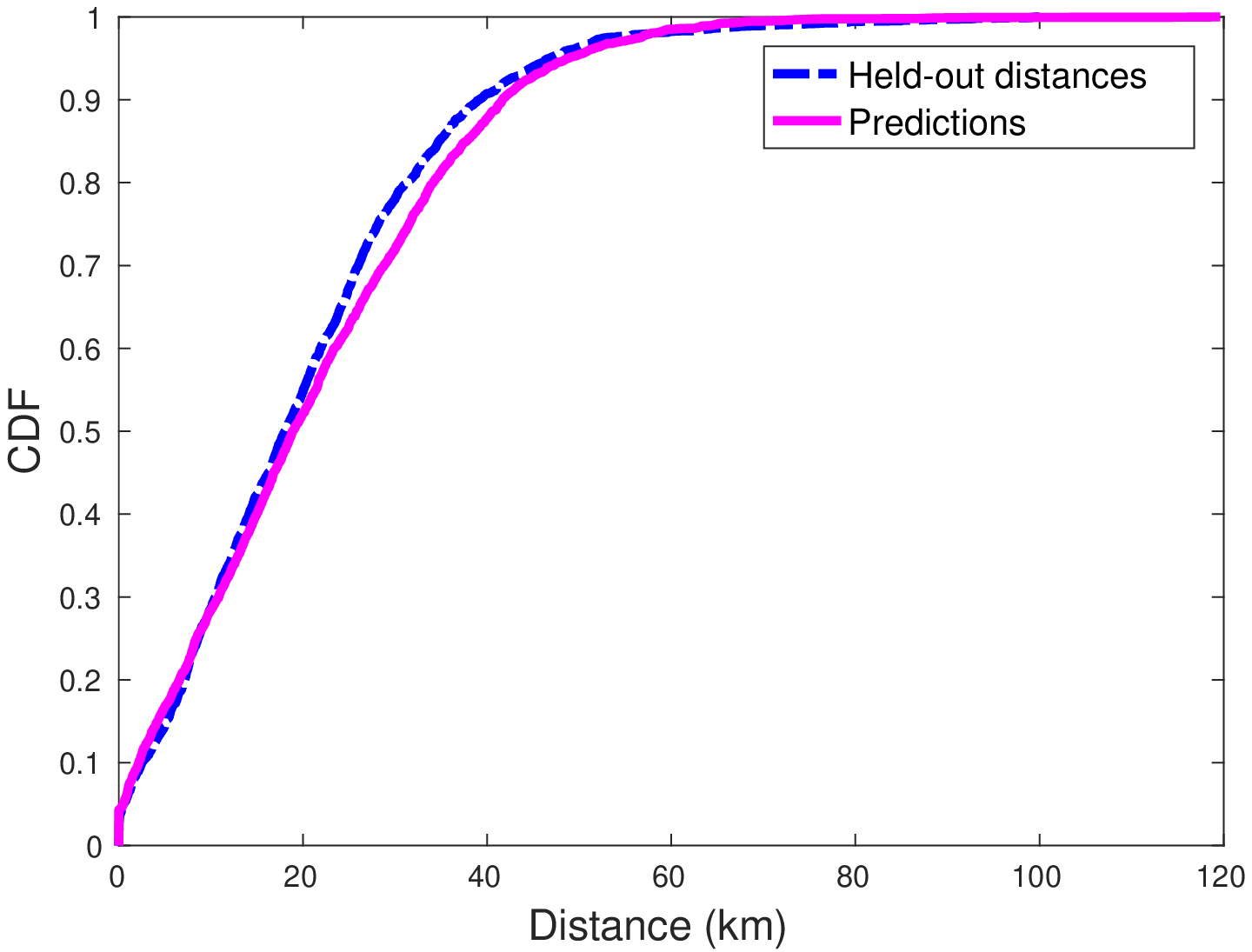,width=5.7cm}}
\end{minipage} 
\caption{\label{fig:holdOutCDF}{{\bf CDF of average relative prediction error (hold-out distances vs. predicted distances) for 100 (left), 1000 (center) and 3668 (right) clients.}}}
\end{centering}
\end{figure*}

%% \subsubsection{Self-consistency checks}
{\bf Self-consistency checks.}  In this analysis, we randomly hold out available OWD values from matrix $X$ and compare them against the predicted values.
%%  by TimeWeaver.  
Figure~\ref{fig:holdOut} shows the distances held-out and the corresponding distances predicted our algorithm. The corresponding CDF of average relative prediction errors are shown in Figure~\ref{fig:holdOutCDF}. For all these different held-out client groups, our approach produced highly accurate estimates of OWDs with an average error rate on the order of less than 2\%. 

\subsection{Applicability to non-US regions} 

The gains that we see in TimeWeaver's ability to make accurate distance estimations are based fundamentally on our ability to accurately filter and extract OWD measurements from NTP logs. Hence, even for other non-US regions that are susceptible to circuitous routing~\cite{gupta2014peering}, we argue that our methods are resistant to routing changes and path asymmetry as long as accurate {\em minimum} OWD measurement is available---which is the {\em only} requirement for TimeWeaver---despite the presence of other large and varying OWD measurements. We plan to investigate this in detail as part of future work.

%% file: applications.tex
%%%% \vspace{-0.2cm}
In this section, we discuss three applications that could benefit from OWD measurements derived from TimeWeaver's different precision tiers, leaving implementation details for future work. We envision either a web services API ({\em e.g.}, BGPmon~\cite{bgpmon}) or a stand-alone service ({\em e.g.}, IDMaps~\cite{Francis2001}) to disseminate OWD data from TimeWeaver to all the applications discussed below as well as many others that are possible.
%% can utilize the opportunistic measurements from NTP servers but leave the implementation details for future work.

{\bf IP geolocation.} IP geolocation has important implications for many applications in the Internet including on-line advertising, content localization and digital rights management. 
%% The common requirement among these applications is the availability of absolute location of IP addresses in terms of longitude/latitude coordinates via a geolocation service ({\em e.g.}, MaxMind~\cite{maxmind}). 
We believe that TimeWeaver-derived OWD measurements are well-suited to form the basis of a new geolocation service. In our initial analysis, we assume highly-accurate minOWD estimates and an embedding in Euclidean space and propose an iterative trilateration-based approach for location estimates~\cite{trilateration}.
We consider tier 3 OWDs to be an ideal choice for use, but it may be feasible to use lower tier measurements depending on the specific requirements of a consuming application. Our main idea is to use {\em geographical closeness}, which is determined using the Vincenty formula~\cite{vincenty}, and {\em small OWDs} between clients and NTP servers as signals for geo-proximity.

Our proposed trilateration-based approach consists of two iterations. First, the algorithm proceeds by creating discs around every NTP server and assigning locations of hosts to be the same as the NTP server when hosts are located within a given disc. The radii of discs ($r$) around NTP servers determines the minimum error in our system.  Second, for every client $c$ that synchronizes with server $s$, we check if the minOWD between $s$ and $c$ are within the disc around $s$. If they are, we assume such clients to be geographically proximal to $s$ and assign the location coordinates as that of $s$. These clients offer opportunities to quantify the geolocation errors as informed by Katz-Bassett {\em et al.}~\cite{Katz06} when the estimates are greater than $r$ from $s$.  The resulting geolocation estimates based on TimeWeaver-derived OWDs would be complementary to the ones produced by prior active measurement-based and commercial efforts ({\em e.g.}, ~\cite{Gueye06,Wong07,maxmind}).
%% including Constraint-Based Geolocation (CBG)~\cite{Gueye06} and Octant~\cite{Wong07}, and the commercial MaxMind IP geolocation database~\cite{maxmind}.

{\bf Census and survey.} Cataloging the {\em active} addresses in the Internet, commonly known as census and survey, has been of interest to the measurement community over the last decade. Notable techniques include active ({\em e.g.}, IPv4 address space using ICMP probes~\cite{heidemann2008census}, using botnet scans~\cite{carna}), passive ({\em e.g.}, IPv4 and IPv6 addresses from BGP updates~\cite{potaroov4, potaroov6}, from traffic captures~\cite{dainotti2013estimating}) and hybrid~\cite{zander2014capturing} measurement techniques. We believe that our approach is complementary to all the aforementioned techniques and offers an opportunity to directly record the active addresses on a daily basis from NTP servers. 

The problem of address visibility at vantage points is first explored by Dainotti {\em et al.}~\cite{dainotti2013estimating}. In addition, for many clients---especially the ones behind a NAT---coverage can be challenging when we measure usage via active probe-based techniques. We posit that the clients observed through TimeWeaver measurements, on an ongoing basis, can effectively address the visibility problem. Furthermore, the ubiquity of NTP usage in the Internet and the fact that clients initiate time synchronization,
as evident from raw numbers (see Table~\ref{tab:statsBasic}) and geographic diversity (see Table~\ref{tab:reach}), can help overcome the coverage issue inherent in active probe-based techniques.  Moreover, latency measures from upper tiers (2--3) could be used to augment information traditionally collected through surveys.

{\bf Network operations and management.} One of the key operational objectives of service providers is to offer highly-available and performant services to their customers. Achieving this objective depends on several factors including robust equipment and services and near real-time tools for operations and management, among others. We believe that shifting the focus from an offline to online analysis to NTP traffic offers the  opportunity to create tools for network operations and management, with key emphasis on monitoring the health of critical Internet services and systems ({\em e.g.}, Akamai's real-time web monitor~\cite{akamaiMonitor}). Furthermore, the pervasive nature of NTP deployments in the Internet offers the ability to monitor {\em (1)} the reachability of critical systems/services using data from all tiers ({\em e.g.}, availability), {\em (2)} performance metrics ({\em e.g.}, rtt, server-to-client and client-to-server OWDs, path asymmetry) using measures from tiers 1--3, and {\em (3)} fault management ({\em e.g.}, outage detection and reporting) using all tiers.

%% file: related.tex
\vspace{-0.1cm}
{\bf Internet path characteristics.} Empirical measurement of Internet path properties ({\em e.g.}, latency, loss, etc.) has been an active area of research for many years.  Early studies of path latencies include Mills's report on RTTs collected using ICMP echo requests~\cite{rfc889} and Bolot's work, which also examined packet loss and queuing~\cite{Bolot93}.
%%Measuring path latencies (typically RTTs) has been studied since the early days of the Internet, for example, when Mills used ICMP echo requests emitted by Fuzzballs to assess latency characteristics~\cite{rfc889}.  Somewhat later, Bolot's study included characterization of round-trip latency, as well as packet loss and queuing~\cite{Bolot93}.
%% After the commercialization of the Internet, it became much harder to fully characterize {\em any} empirical property of the Internet, including latency. 
The landmark work of Paxson used a set of specially-deployed systems to measure packet loss and latency~\cite{paxsonthesis}, and has informed much of the ongoing work in this area.  There are a number of efforts today that take a similar approach of having specially-deployed systems to collect an essentially continuous stream of measurements such as latency, loss, and routing~\cite{ark,pinger,ripeatlas}.
%% surveyor used commodity PCs with GPS cards for one-way delays~\cite{}

There are a number of specific efforts that have focused on accurate estimation of OWD.  A common approach is to assume host clocks have been synchronized ({\em e.g.},~\cite{Hernandez07,Devito08,Shin11}) and to accept OWD measurements at face value. Other work has explicitly addressed correction for clock offset, drift, and skew, {\em e.g.}, ~\cite{Pasztor01,Pasztor02,paxsonthesis}.  Yet other works have attempted to estimate OWD using timestamps in flow records~\cite{Kogel11} or through analysis of multiple one-way measurements collected from a group of unsynchronized hosts~\cite{Gurewitz06,Vakili12}.
An extensive analysis of delay and path asymmetry was done by Pathak {\em et al.}~\cite{Pathak08} using the \texttt{owping} tool~\cite{rfc4656,rfc2679}, which 
%% program to measure OWD, which is an implementation of the one-way active measurement protocol~\cite{rfc4656,rfc2679}.  \texttt{owping} 
estimates OWD by obtaining clock drift information from the local NTP daemon and correcting for it at the receiver. Using data collected over 10 days using the Planetlab infrastructure, Pathak {\em et al.} analyzed OWD and path-level asymmetries, finding some paths in their study to exhibit significant and dynamic asymmetric qualities.

Our work generalizes and extends prior work that identifies NTP as a source of latency measurement~\cite{ntpFilter, Ridoux10, Durairajan2015HotNets}.  In particular, we develop a comprehensive filtering algorithm based on detailed examination of the NTP codebase that enables accurate OWDs to be identified, and we develop and propose new methods for distance estimation and other applications that are based on the availability of these measurements.

{\bf Internet distance estimation.} Apart from measuring latencies, there have been a variety of techniques developed to estimate latencies between arbitrary nodes in the Internet. IDMaps~\cite{Francis2001} examined network distance prediction from a topological perspective and influenced later work on King~\cite{gummadi2002}, which expands on the IDMaps technique but uses DNS servers as landmarks, and Meridian~\cite{Wong2005} which probes landmarks on demand to predict network distances.
%% , which performs on-demand probing from  carefully chosen landmarks to determine closest landmark location and then predicts the network distances. 
The work by Ng and Zhang on GNP~\cite{ng2002} uses a low-dimensional Euclidean space to embed the nodes by relying on well-known pivots (or landmarks). Similar to GNP, Lighthouse~\cite{pias2003} uses a transition matrix to achieve embedding with reference to any pivots. Tang {\em et al.} propose a virtual landmark-based embedding scheme~\cite{tang2003} which is computationally efficient and is independent of landmark positions. Subsequent efforts used different embedding systems, resulting in different performance and accuracy characteristics~\cite{dabek2004, shavitt2004}, and the work by Mao {\em et al.}~\cite{mao2004} proposes matrix factorization techniques to determine network distances. One of the interesting questions raised by Madhyastha {\em et al.}~\cite{madhyastha2006lat} regarding matrix factorization is how OWDs from landmarks to arbitrary clients might be measured. In light of our work, NTP servers naturally become the landmarks and the OWDs to a large and distributed set of clients are easily obtained.

%{\bf IP geolocation.} Identifying the geographic location of fixed (non-mobile) Internet hosts has received significant attention in prior work, {\em e.g.,}~\cite{Katz06, Wang11, Wong07, PS01, Gueye06, Eriksson12, Rasti10, Feldman12, Lee2016}.  Most prior work uses traceroute probes to assess topological and delay characteristics between nodes with known locations and other hosts, and some work also uses ``unDNS'' techniques to reverse engineer locations of nodes based on DNS naming conventions.  While we also use latency as a proxy for distance, we rely on having accurate OWDs, which has not been possible with prior work due to the nature of the latency measurement techniques used ({\em i.e.}, both ping and traceroute provide RTT measurements, not OWD).  While we use DNS naming hints for validation of our geolocation estimates, our methods otherwise do not require them.

%% file: summary.tex
%%%% \vspace{-0.2cm}
Accurate one-way delay measurements between hosts are important for many Internet applications and protocols, as well for effective monitoring and measurement-based network analysis.  In this paper, we consider the problem of gathering OWD measurements in the Internet, at scale.  Our approach is based on passive measurement and analysis of the timestamps and other information that can be gleaned through traces of NTP traffic. Based on detailed analysis of the protocol, of the codebase, of NTP traces, and of laboratory-based experiments, we develop a new method and tool called TimeWeaver for correcting and filtering OWD measurements extracted from NTP packets.  Critically, TimeWeaver leverages NTP's polling interval and heuristics similar to those used in the NTP code to identify samples that indicate strong synchronization between a client and server, as well as samples that indicate poor synchronization.  The resulting OWD estimates are classified in a precision tier, which provides an explicit context for understanding the accuracy and potential utility of the measurements.

We apply TimeWeaver to a $\sim$1TB corpus of NTP trace data collected over a period of 30 days from 19 servers in the US.  The data represent synchronization activity from hundreds of millions of broadly distributed clients.  We compare the resulting OWD estimates to probe-based RTT samples as well as to a prior NTP filtering method.  We find that sum of forward and reverse OWD measurements correlates well with RTT measurements and that TimeWeaver offers much greater accuracy for its resulting top-tier OWD estimates than prior work.  We also find that TimeWeaver's filtering and precision classifying approach results in a much broader set of OWD measurements that can be extracted from raw NTP traffic.  When we analyze the distributional characteristics of TimeWeaver's OWD measurements, we find that lower precision tiers exhibit a broad range of OWDs while the top precision tier exhibits a fairly narrow range of OWD values. % We also examine asymmetry in OWD measurements and find that, similar to prior work, delays can be highly asymmetric, thus implying that RTT/2 is an unreliable proxy for OWD.

To illustrate the utility of having accurate OWD measurements, we approach the problem of distance estimation under the assumption that accurate minOWD data is available and that measurement data can be missing or incomplete.  We use minOWD estimates from a subset of NTP clients for which we have the highest tier (tier 3) estimates, and which contact multiple servers. We apply iterative hard-thresholded SVD to complete the matrix of inter-host delays.  We find that the resulting estimates are highly accurate with relative errors are on the order of 2\%.  
%% We also show that when a sizable number of measurements are missing, our approach reduces distance estimation error by a factor of 4, on average, compared to prior methods.

Currently, we are extending TimeWeaver for real-time OWD estimation. This requires a source for real-time NTP measurements at strategic locations ({\tt pool.ntp.org} is a simple and cost effective option), and adaptation of TimeWeaver to operate on streams of NTP data.

Finally, to outline the utility of having a large corpus of OWD measurements of varying precision and accuracy using TimeWeaver, we also discuss three application areas: IP geolocation, census and survey of active Internet addresses, and network operations and management.  Our proposed IP geolocation algorithm is based on having accurate minOWD measurements, as could obtained from TimeWeaver's tier 3 measurements, and addresses the problem using an iterative trilateration-based approach.  We propose to use OWD estimates from multiple precision tiers in active address survey and census, and as a way to provide a broad view of address usage over time due to NTP's ubiquity.  Finally, we believe that TimeWeaver-based OWD measurements from various tiers have significant potential for continuous assessment of network availability, performance and fault monitoring.  We intend to consider each of these areas in our future work.

% Under the assumption of line-of-sight approximation of minOWD, we address the problem of IP geolocation using an iterative trilateration-based approach. We show that our resulting location estimates are as precise as Octant and 5x more accurate than CBG. We find that locations of 90\% of the clients can be predicted within 8km from the closest NTP server when compared against predictions using the MaxMind commercial geolocation database and estimates for clients that have location hints are twice as accurate as those using the MaxMind.

% There are a number of ways in which this work can be expanded.  Our current focus is on developing the capability to include clients with accurate OWD estimates which only contact a single server.  We are also evaluating how accurate OWD data can be extracted from NTP to create a distance estimation service.  OWD data can also be used in a variety of empirical analyses including path dynamics.  Finally, our insights on NTP behavior can also lead to improvements on time synchronization.

%% file: paper.bbl
\begin{thebibliography}{10}
\providecommand{\url}[1]{#1}
\csname url@samestyle\endcsname
\providecommand{\newblock}{\relax}
\providecommand{\bibinfo}[2]{#2}
\providecommand{\BIBentrySTDinterwordspacing}{\spaceskip=0pt\relax}
\providecommand{\BIBentryALTinterwordstretchfactor}{4}
\providecommand{\BIBentryALTinterwordspacing}{\spaceskip=\fontdimen2\font plus
\BIBentryALTinterwordstretchfactor\fontdimen3\font minus
  \fontdimen4\font\relax}
\providecommand{\BIBforeignlanguage}[2]{{%
\expandafter\ifx\csname l@#1\endcsname\relax
\typeout{** WARNING: IEEEtran.bst: No hyphenation pattern has been}%
\typeout{** loaded for the language `#1'. Using the pattern for}%
\typeout{** the default language instead.}%
\else
\language=\csname l@#1\endcsname
\fi
#2}}
\providecommand{\BIBdecl}{\relax}
\BIBdecl

\bibitem{dabek2004}
{F. Dabek and R. Cox and F. Kaashoek and R. Morris}, ``{Vivaldi: A
  Decentralized Network Coordinate System},'' in \emph{ACM SIGCOMM}, 2004.

\bibitem{mao2006ides}
{Y. Mao and L. K. Saul and J. M. Smith}, ``{IDES: An Internet Distance
  Estimation Service for Large networks},'' in \emph{IEEE JSAC}, 2006.

\bibitem{chen2009phoenix}
Y.~Chen, X.~Wang, X.~Song, E.~Lua, C.~Shi, X.~Zhao, B.~Deng, and X.~Li,
  ``{Phoenix: Towards an Accurate, Practical and Decentralized Network
  Coordinate System},'' in \emph{Networking}, 2009.

\bibitem{liao2010network}
{Y. Liao and P. Geurts and G. Leduc}, ``{Network Distance Prediction based on
  Decentralized Matrix Factorization},'' in \emph{Networking}, 2010.

\bibitem{jain2002end}
M.~Jain and C.~Dovrolis, ``{End-to-end Available Bandwidth: Measurement
  Methodology, Dynamics, and Relation with TCP Throughput},'' in \emph{ACM
  SIGCOMM}, 2002.

\bibitem{sommers2006proposed}
J.~Sommers, P.~Barford, and W.~Willinger, ``{A Proposed Framework for
  Calibration of Available Bandwidth Estimation Tools},'' in \emph{IEEE ISCC},
  2006.

\bibitem{Gueye06}
B.~Gueye, A.~Ziviani, M.~Crovella, and S.~Fdida, ``{Constraint-Based
  Geolocation of Internet Hosts},'' \emph{{IEEE/ACM TON}}, 2006.

\bibitem{Wong07}
B.~Wong, I.~Stoyanov, and E.~Sirer, ``{Octant: A Comprehensive Framework for
  the Geolocation of Internet Hosts},'' in \emph{USENIX NSDI}, 2007.

\bibitem{krishnan2009moving}
R.~Krishnan, H.~V. Madhyastha, S.~Srinivasan, S.~Jain, A.~Krishnamurthy,
  T.~Anderson, and J.~Gao, ``{Moving Beyond End-to-end Path Information to
  Optimize CDN Performance},'' in \emph{ACM IMC}, 2009.

\bibitem{rowstron2001pastry}
A.~Rowstron and P.~Druschel, ``{Pastry: Scalable, Decentralized Object
  Location, and Routing for Large-scale Peer-to-peer Systems},'' in
  \emph{IFIP/ACM Middleware}, 2001.

\bibitem{paxsonthesis}
V.~Paxson, ``{Measurements and Analysis of End-to-end Internet Dynamics},''
  Ph.D. dissertation, University of California, Berkeley, 1997.

\bibitem{sanchez13}
M.~S{\'a}nchez, J.~Otto, Z.~Bischof, D.~Choffnes, F.~Bustamante,
  B.~Krishnamurthy, and W.~Willinger, ``{Dasu: Pushing Experiments to the
  Internet's Edge.}'' in \emph{Usenix NSDI}, 2013.

\bibitem{zpds01}
Y.~Zhang and N.~Duffield, ``{On the Constancy of Internet Path Properties},''
  in \emph{ACM IMW}, 2001.

\bibitem{Durairajan2015HotNets}
{R. Durairajan and S. Mani and J. Sommers and P. Barford}, ``{Time's Forgotten:
  Using NTP to Understand Internet Latency},'' in \emph{ACM HotNets}, 2015.

\bibitem{ntpCodeBase}
``{NTP Codebase.}'' \url{https://github.com/ntp-project/ntp}, 2016.

\bibitem{Francis2001}
P.~Francis, S.~Jamin, C.~Jin, Y.~Jin, D.~Raz, Y.~Shavitt, and L.~Zhang,
  ``{IDMaps: A Global Internet Host Distance Estimation Service},''
  \emph{IEEE/ACM TON}, 2001.

\bibitem{ng2002}
E.~Ng and H.~Zhang, ``{Predicting Internet Network Distance with
  Coordinates-based Approaches},'' in \emph{IEEE INFOCOM}, 2002.

\bibitem{chunikhina2014}
{E. Chunikhina and R. Raich and T. Nguyen}, ``{Performance Analysis for Matrix
  Completion via Iterative Hard-thresholded SVD},'' in \emph{IEEE SSP
  Workshop}, 2014.

\bibitem{rfc958}
D.~Mills, ``{Network Time Protocol (NTP)},''
  \url{https://tools.ietf.org/html/rfc958}, September 1985.

\bibitem{rfc5905}
D.~Mills, J.~Martin, J.~Burbank, and W.~Kasch, ``{Network Time Protocol Version
  4: Protocol and Algorithms Specification},''
  \url{https://tools.ietf.org/html/rfc5905}, June 2010.

\bibitem{rfc1305}
D.~Mills, ``{Network Time Protocol (Version 3): Specification, Implementation
  and Analysis},'' \url{https://www.ietf.org/rfc/rfc1305.txt}.

\bibitem{ntpClockDiscipline}
``{NTP Clock Discipline Algorithm.}''
  \url{https://www.eecis.udel.edu/~mills/ntp/html/discipline.html}.

\bibitem{nptfaq}
``{NTP Polling Interval.}''
  \url{http://www.eecis.udel.edu/~mills/ntp/html/poll.html}.

\bibitem{rfc1769}
D.~Mills, ``{Simple Network Time Protocol (SNTP)},''
  \url{https://tools.ietf.org/html/rfc1769}, March 1995.

\bibitem{sathiya2016}
S.~K. Mani, R.~Durairajan, P.~Barford, and J.~Sommers, ``{MNTP:} enhancing time
  synchronization for mobile devices,'' in \emph{ACM IMC}, 2016.

\bibitem{ntppool}
``{NTP Pool Servers},'' \url{http://pool.ntp.org}.

\bibitem{ntpFilter}
``{NTP Clock Filter Algorithm.}''
  \url{https://www.eecis.udel.edu/~mills/ntp/html/filter.html}.

\bibitem{Ridoux10}
J.~Ridoux and D.~Veitch, ``{Principles of Robust Timing over the Internet},''
  \emph{Queue}, 2010.

\bibitem{Paxson04}
V.~Paxson, ``{Strategies for Sound Internet Measurement},'' in \emph{ACM IMC},
  2004.

\bibitem{Madhyastha2006}
H.~V. Madhyastha, T.~Isdal, M.~Piatek, C.~Dixon, T.~Anderson, A.~Krishnamurthy,
  and A.~Venkataramani, ``{iPlane: An Information Plane for Distributed
  Services},'' in \emph{ACM SOSP}, 2006.

\bibitem{Katz-Bassett2008}
E.~Katz-Bassett, H.~Madhyastha, J.~John, A.~Krishnamurthy, D.~Wetherall, and
  T.~Anderson, ``{Studying Black Holes in the Internet with Hubble},'' in
  \emph{USENIX NSDI}, 2008.

\bibitem{ripeatlas}
``{RIPE Atlas},'' \url{https://atlas.ripe.net}, 2015.

\bibitem{mailPhilip}
``{Timeout on Ping Measurements},''
  \url{https://www.ripe.net/ripe/mail/archives/ripe-atlas/2013-July/000891.html}.

\bibitem{source194}
``{How Many Countries Are In The World?}''
  \url{http://www.worldatlas.com/nations.htm}.

\bibitem{source195}
``{How Many Countries Are In The World?}''
  \url{http://www.worldometers.info/geography/how-many-countries-are-there-in-the-world/}.

\bibitem{source196}
``{How Many Countries?}'' \url{http://www.infoplease.com/ipa/A0932875.html}.

\bibitem{source247}
``How many countries are there in the world?''
  \url{http://www.world-country.com/}.

\bibitem{maxMindSource}
``Source code of geoip.c used by maxmind.''
  \url{https://searchcode.com/file/18446750/GeoIP-1.4.8/libGeoIP/GeoIP.c}.

\bibitem{ieeeptp}
{IEEE}, ``{IEEE 1588 Precision Time Protocol (PTP), Version 2 Specification},''
  March 2008.

\bibitem{millsPTP}
{D.L. Mills}, ``{IEEE 1588 Precision Time Protocol (PTP), Version 2
  Specification},'' \url{https://www.eecis.udel.edu/~mills/ptp.html}.

\bibitem{geodetic}
``{World Geodetic System 1984},''
  \url{http://www.dtic.mil/docs/citations/ADA167570}.

\bibitem{vincenty}
``{Vincenty Formula.}''
  \url{https://en.wikipedia.org/wiki/Vincenty\%27s_formulae}.

\bibitem{singla2014}
{A. Singla and B. Chandrasekaran and B. Godfrey and B. Maggs}, ``{The Internet
  at the Speed of Light},'' in \emph{ACM HotNets}, 2014.

\bibitem{cymru}
``{Team Cymru whois lookup service.}''
  \url{http://www.team-cymru.org/IP-ASN-mapping.html\#whois}.

\bibitem{gupta2014peering}
A.~Gupta, M.~Calder, N.~Feamster, M.~Chetty, E.~Calandro, and E.~Katz-Bassett,
  ``{Peering at the Internet's Frontier: A First Look at ISP Interconnectivity
  in Africa},'' in \emph{PAM}, 2014.

\bibitem{bgpmon}
``{BGPmon Web Services API.}''
  \url{http://bgpmon.net/bgpmon-web-services-api/}.

\bibitem{trilateration}
``{Trilateration formulation.}''
  \url{https://en.wikipedia.org/wiki/Trilateration\#Derivation}.

\bibitem{Katz06}
{E. Katz-Bassett and J. P. John and A. Krishnamurthy and D. Wetherall and T.
  Anderson and Y. Chawathe}, ``{Towards IP Geolocation Using Delay and Topology
  Measurements},'' in \emph{ACM IMC}, 2006.

\bibitem{maxmind}
``{MaxMind: IP Geolocation and Online Fraud Prevention},''
  \url{https://www.maxmind.com/}, 2015.

\bibitem{heidemann2008census}
J.~Heidemann, Y.~Pradkin, R.~Govindan, C.~Papadopoulos, G.~Bartlett, and
  J.~Bannister, ``{Census and Survey of the Visible Internet},'' in \emph{ACM
  IMC}, 2008.

\bibitem{carna}
``{Internet Census 2012.}''
  \url{https://internetcensus2012.bitbucket.io/paper.html}.

\bibitem{potaroov4}
``{IPv4 reports from bgp.potaroo.net.}''
  \url{http://bgp.potaroo.net/index-ale.html}.

\bibitem{potaroov6}
``{IPv6 reports from bgp.potaroo.net.}''
  \url{http://bgp.potaroo.net/index-v6.html}.

\bibitem{dainotti2013estimating}
A.~Dainotti, K.~Benson, A.~King, M.~Kallitsis, E.~Glatz, X.~Dimitropoulos
  \emph{et~al.}, ``{Estimating Internet Address Space Usage through Passive
  Measurements},'' \emph{ACM SIGCOMM CCR}, 2013.

\bibitem{zander2014capturing}
S.~Zander, L.~L. Andrew, and G.~Armitage, ``{Capturing Ghosts: Predicting the
  Used IPv4 Space by Inferring Unobserved Addresses},'' in \emph{ACM IMC},
  2014.

\bibitem{akamaiMonitor}
``{Akamai Real-time Web Monitor.}''
  \url{https://www.akamai.com/us/en/solutions/intelligent-platform/visualizing-akamai/real-time-web-monitor.jsp}.

\bibitem{rfc889}
D.~Mills, ``{Internet Delay Experiments},''
  \url{https://tools.ietf.org/html/rfc889}, December 1983.

\bibitem{Bolot93}
J.~Bolot, ``{End-to-end Packet Delay and Loss Behavior in the Internet},'' in
  \emph{SIGCOMM CCR}, 1993.

\bibitem{ark}
``{CAIDA's Ark Project.}'' \url{http://www.caida.org}.

\bibitem{pinger}
``{PingER: Ping End-to-end Reporting},''
  \url{http://www-iepm.slac.stanford.edu/pinger/}, 2015.

\bibitem{Hernandez07}
A.~Hernandez and E.~Magana, ``{One-way Delay Measurement and
  Characterization},'' in \emph{IEEE ICNS}, 2007.

\bibitem{Devito08}
L.~D. Vito, S.~Rapuano, and L.~Tomaciello, ``{One-way Delay Measurement: State
  of the Art},'' \emph{IEEE TIM}, 2008.

\bibitem{Shin11}
M.~Shin, M.~Park, D.~Oh, B.~Kim, and J.~Lee, ``{Clock Synchronization for
  One-way Delay Measurement: A Survey},'' in \emph{Advanced Communication and
  Networking}, 2011.

\bibitem{Pasztor01}
A.~Pasztor and D.~Veitch, ``{A Precision Infrastructure for Active Probing},''
  in \emph{PAM}, 2001.

\bibitem{Pasztor02}
------, ``{PC-based Precision Timing Without GPS},'' in \emph{ACM SIGMETRICS},
  2002.

\bibitem{Kogel11}
J.~K{\"o}gel, ``{One-way Delay Measurement Based on Flow Data: Quantification
  and Compensation of Errors by Exporter Profiling},'' in \emph{ICOIN}, 2011.

\bibitem{Gurewitz06}
O.~Gurewitz, I.~Cidon, and M.~Sidi, ``{One-way Delay Estimation using
  Network-wide Measurements},'' \emph{IEEE/ACM TON}, vol.~14, no.~SI, 2006.

\bibitem{Vakili12}
A.~Vakili and J.-C. Gregoire, ``{Accurate One-way Delay Estimation: Limitations
  and Improvements},'' \emph{IEEE TIM}, 2012.

\bibitem{Pathak08}
A.~Pathak, H.~Pucha, Y.~Zhang, Y.~C. Hu, and Z.~M. Mao, ``{A Measurement Study
  of Internet Delay Asymmetry},'' in \emph{PAM}, 2008.

\bibitem{rfc4656}
S.~Shalunov, B.~Teitelbaum, A.~Karp, J.~Boote, and M.~Zekauskas, ``{RFC 4656: A
  One-way Active Measurement Protocol (OWAMP)},'' September 2006.

\bibitem{rfc2679}
G.~Almes, S.~Kalidindi, and M.~Zekauskas, ``{RFC 2679: A One-way Delay Metric
  for IPPM},'' September 1999.

\bibitem{gummadi2002}
K.~Gummadi, S.~Saroiu, and S.~Gribble, ``{King: Estimating Latency between
  Arbitrary Internet End Hosts},'' in \emph{ACM IMW}, 2002.

\bibitem{Wong2005}
B.~Wong, A.~Slivkins, and E.~Sirer, ``{Meridian: A Lightweight Network Location
  Service Without Virtual Coordinates},'' in \emph{ACM SIGCOMM}, 2005.

\bibitem{pias2003}
M.~Pias, J.~Crowcroft, S.~Wilbur, T.~Harris, and S.~Bhatti, ``{Lighthouses for
  Scalable Distributed Location},'' in \emph{USENIX IPTPS}, 2003.

\bibitem{tang2003}
L.~Tang and M.~Crovella, ``{Virtual Landmarks for the Internet},'' in \emph{ACM
  IMC}, 2003.

\bibitem{shavitt2004}
Y.~Shavitt and T.~Tankel, ``{On the Curvature of the Internet and its Usage for
  Overlay Construction and Distance Estimation},'' in \emph{IEEE INFOCOM},
  2004.

\bibitem{mao2004}
Y.~Mao and L.~K. Saul, ``{Modeling Distances in Large-scale Networks by Matrix
  Factorization},'' in \emph{ACM IMC}, 2004.

\bibitem{madhyastha2006lat}
H.~V. Madhyastha, T.~Anderson, A.~Krishnamurthy, N.~Spring, and
  A.~Venkataramani, ``{A Structural Approach to Latency Prediction},'' in
  \emph{ACM IMC}, 2006.

\end{thebibliography}
